\documentclass[a4paper,fleqn,usenatbib,onecolumn]{mnras}
\usepackage[T1]{fontenc}
\usepackage[dvipsnames]{xcolor}
\usepackage{ae,aecompl}
\usepackage{graphicx}	% Including figure files
\usepackage{amsmath}	% Advanced maths commands
\usepackage{amssymb}	% Extra maths symbols
\usepackage{booktabs}
\usepackage{stackengine}
\usepackage{multirow}
\newcommand{\be}{\begin{equation}}
\newcommand{\ee}{\end{equation}}
\newcommand{\bq}{\begin{eqnarray}}
\newcommand{\eq}{\end{eqnarray}}
\title[Watching the Universe's acceleration era]{Watching the Universe's acceleration era with the SKAO}

\author[C. M. J. Marques et al.]{
C. M. J. Marques,$^{1,2,3}$\thanks{E-mail: Catarina.Marques@astro.up.pt (CMJM)}
C. J. A. P. Martins,$^{1,3}$\thanks{E-mail: Carlos.Martins@astro.up.pt (CJAPM)}
and B. Gilabert L\'opez$^{4}$\thanks{E-mail: blgilabl14@alumnes.ub.edu (BGL)}
\\
$^{1}$Centro de Astrof\'{\i}sica da Universidade do Porto, Rua das Estrelas, 4150-762 Porto, Portugal\\
$^{2}$Faculdade de Ci\^encias, Universidade do Porto, Rua do Campo Alegre, 4150-007 Porto, Portugal\\
$^{3}$Instituto de Astrof\'{\i}sica e Ci\^encias do Espa\c co, CAUP, Rua das Estrelas, 4150-762 Porto, Portugal\\
$^{4}$Faculty of Physics, University of Barcelona, Mart\'{\i} i Franqu\`es 1-11, 08028 Barcelona, Spain
}

\date{Accepted XXX. Received YYY; in original form ZZZ}
\pubyear{2023}
\begin{document}
\label{firstpage}
\pagerange{\pageref{firstpage}--\pageref{lastpage}}
\maketitle

\begin{abstract}
The possibility of watching the Universe expand in real time and in a model-independent way, first envisaged by Allan Sandage more than 60 years ago and known as the redshift drift, is within reach of forthcoming astrophysical facilities, particularly the Extremely Large Telescope (ELT) and the Square Kilometre Array Observatory (SKAO). The latter, probing lower redshifts, enables us to watch the Universe's acceleration era in real time, while the former does the same for the matter era. We use Fisher Matrix Analysis techniques, which we show to give comparable results to those of a Markov Chain Monte Carlo approach, to discuss forecasts for SKAO measurements of the redshift drift and their cosmological impact. We consider specific fiducial cosmological models but mainly rely on a more agnostic cosmographic series (which includes the deceleration and jerk parameters), and we also discuss prospects for measurements of the drift of the drift. Overall, our analysis shows that SKAO measurements, with a reasonable amount of observing time, can provide a competitive probe of the low-redshift accelerating Universe.
\end{abstract}


%%%%%%%%%%%%%%%%% BODY OF PAPER %%%%%%%%%%%%%%%%%%

\section{Introduction}

Identifying the physical mechanism responsible for the recent accelerated expansion of the Universe is a compelling challenge of modern observational cosmology, not least because this acceleration highlights the presence of hitherto unknown physics. Either this is due to a cosmological constant whose value must be many orders of magnitude smaller than current quantum field theory expectations, there are additional cosmologically relevant dynamical degrees of freedom, or some other large-scale modification of locally know gravitational laws. It is clear that our current standard model, $\Lambda$CDM, can be no more than a reasonably good phenomenological approximation to an underlying but currently undiscovered model.

Arguably the main bottleneck is this quest is the fact that our traditional cosmological observations are model-dependent, and their analysis is thereby limited by our prior knowledge of plausible models. Nevertheless, cosmology holds a conceptually simple (though operationally challenging) observable which is model-independent (e.g. it is a measurement done in velocity space, which relies on no assumptions on gravity, geometry or clustering): the redshift drift of objects following the cosmological expansion, also known as the Sandage Test  \citep[][note that the latter reference is an appendix to, and
immediately follows, the former]{Sandage,Mcvittie}. Although the idea was occasionally considered in the subsequent decades \citep{Lake1,Loeb,Lake2}, the practical challenge stems from the fact that the characteristic cosmological time scales involved are orders of magnitude larger than human time scales and, therefore, redshift drift measurements require an exquisite sensitivity. Specifically, current facilities allow only sensitivities around three orders of magnitude worse than the expected signal for $\Lambda$CDM model \citep{Darling,Cooke}, and it is clear that both of these are dominated by systematic uncertainties.

Two forthcoming astrophysical facilities are expected to provide the first detection of the redshift drift signal\footnote{More recently, a less direct method relying on time delays in strong gravitational lenses has been suggested \citep{Piattella,Wucknitz,Wang,Helbig}, but the feasibility of this method has not been demonstrated.}. The first is the European Extremely Large Telescope (ELT), and specifically its high-resolution spectrograph \citep{HIRES}, currently known as ArmazoNes high Dispersion Echelle Spectrograph (ANDES) \citep{Marconi}. A first detailed assessment of the feasibility of such measurements was reported by \citet{Liske}, while more recently \citet{Qubrics} provided an optimized `Golden Sample' of targets for these measurements and \citet{Marques} discussed the wider cosmological impact of such ANDES redshift drift measurements, and also publicly released the corresponding forecast code \citep{FRIDDA}. This code is based on Fisher Matrix Analysis techniques \citep{FMA1,FMA2}, and we will also use it in the present work. Since the ELT redshift drift measurements predominantly rely on the Lyman-$\alpha$ forest, the ELT can do these measurements only at redshifts $z\sim2$ and beyond. In this regard, it is worthy of note that redshift $z\sim2$ is significant for this observable, since it is where one expects, at least in the context of the $\Lambda$CDM model, the redshift drift signal to change sign\footnote{Note that this redshift does not coincide with the beginning of the acceleration epoch; this is further discussed e.g. in \citet{Alves}.}. To be  specific, the sign is expected to be positive at lower redshifts and negative at higher redshifts.

The second such astrophysical probe is the Square Kilometre Array Observatory (SKAO), which is the focus of our present work. This is an ideal complement to the ELT, probing lower redshifts (presumably $z<1$) through entirely different observational techniques. It is worthy of note that, although redshift drift measurements at high redshift can have larger downstream effects in terms of constraining cosmological models (since high-redshift observations are comparatively more scarce), a direct measurement of a positive drift signal---{which for observationally plausible models\footnote{We are ignoring unphysical sources of negative energy density.} is possible only at low redshifts---is conceptually important, since it implies a violation of the strong energy condition, and hence the presence of some form of dark energy  (be it a cosmological constant or some alternative mechanism) accelerating the Universe \citep{Liske,Uzan,Quercellini,Heinesen}. Unfortunately, no SKAO redshift drift feasibility studies with a level of detail analogous those for the ELT currently exist. The one available study is reported in \citet{Klockner}, which suggests that the SKAO can (depending on hardware configurations and other uncertain instrumental factors) detect the redshift drift signal with percent-level sensitivity at redshift $z<1$, with a 12 year experiment time and a 0.5 year on-sky measurement time. Under those assumptions the SKAO might go even further and measure the second derivatives of redshift --- that is, the drift of the drift \citep{Second}.

In what follows we take the analysis of \citet{Klockner} at face value and build upon previous works \citep{Second,Alves,Rocha} and the recently released FRIDDA\footnote{Available at \url{https://github.com/CatarinaMMarques/FisherCosmology}.} code \citep{FRIDDA} to provide a more detailed assessment of the cosmological impact of redshift drift measurements by the SKAO. We provide a comparison between Fisher Matrix and Markov Chain Monte Carlo (MCMC) based forecasts and consider the impact of the choice of priors on our results. In addition to considering the redshift drift per se, we also discuss prospects for measurements of higher-order derivatives of the redshift, building upon an earlier analysis in \citet{Second}. Our analysis relies on specific fiducial dark energy model choices, but we mainly express constraints using the more model-independent cosmographic approach \citep{Visser}. Finally, although we mainly focus on the deceleration and jerk parameters, we generically consider higher-order terms in the cosmographic series, and in particular we discuss the impact of the choice of the order at which the cosmographic series is truncated on the derived constraints, and how this choice correlates with the redshift range of the available measurements.

%%%%%%%%%%%%%%%%%%%%%%%%%%%%%%%%%%%%%%%%%%%%%%%%%%%%%%%%%%%%%%%%%%%%%%%%%%%%%%%%%
%%%%%%%%%%%%%%%%%%%%%%%%%%%%%%%%%%%%%%%%%%%%%%%%%%%%%%%%%%%%%%%%%%%%%%%%%%%%%%%%%
\section{Theoretical Formalism and Assumptions}

We start by briefly introducing our conceptual analysis formalism, defining the relevant quantities, and identifying our fiducial model assumptions. This is, to a large extent, a review of previous works, but it is included here to make the current work self-contained. The reader may refer to the cited works for additional details.

\subsection{Standard redshift drift}

The redshift drift of an astrophysical object following the cosmological expansion, for an observer looking at said object over some time span $\Delta t$, can be written \citep{Sandage,Liske}
\be\label{eq1}
\frac{\Delta z}{\Delta t}=H_{\mathrm{0}} \left[1+z-E(z)\right]\,;
\ee
in practice the measurement is done in velocity space, and therefore the directly relevant observable is a spectroscopic velocity
\be\label{specvel}
\Delta v=\frac{c\Delta z}{1+z}=(cH_{\mathrm{0}}\Delta t)\left[1-\frac{E(z)}{1+z}\right]\,.
\ee
For future convenience we have defined the re-scaled Hubble parameter $E(z)=H(z)/H_{\mathrm{0}}$, with $H_{\mathrm{0}}$ denoting the Hubble constant. Naturally the Hubble parameter dependence encodes the model-dependence of the signal. When focusing our analysis on a specific model, our fiducial model will be the Chevallier-Polarski-Linder (CPL) parametrization of dark energy \citep{CPL1,CPL2}, for which the Friedmann equation takes the form
\be\label{cpl}
E^2(z)=\Omega_{\mathrm{k}}(1+z)^2+\Omega_{\mathrm{m}}(1+z)^3+\Omega_\phi(1+z)^{3(1+w_{\mathrm{0}}+w_{\mathrm{a}})}\exp{\left[-\frac{3w_{\mathrm{a}}z}{1+z}\right]}\,.
\ee

In what follows we set the spatial curvature to zero ($\Omega_{\mathrm{k}}=0$), leaving the analysis of its impact to a separate work. The choice of $w_{\mathrm{0}}=-1$ and $w_{\mathrm{a}}=0$ corresponds to $\Lambda$CDM. This, together with $\Omega_{\mathrm{m}}=0.3$ (so $\Omega_\phi=1-\Omega_{\mathrm{m}}=0.7$) and $H_{\mathrm{0}}=70$ km/s/Mpc will be our assumed fiducial model parameter choice, unless otherwise stated. Detailed analyses of of the sensitivity of redshift drift measurements to each of these model parameters \citep{Alves,Esteves} show that the largest sensitivity occurs for the matter density $\Omega_{\mathrm{m}}$, with the sensitivity being a monotonic function of redshift. For $w_{\mathrm{0}}$ and $w_{\mathrm{a}}$ the sensitivity is weaker but, for parameter values near the above fiducial values, it is maximal at redshifts slightly below $z=1$, making the SKAO an ideal facility to probe them by these means.

%%%%%%%%%%%%%%%%%%%%%%%%%%%%%%%%%%%%%%%%%%%%%%%%%%%%%%%%%%%%%%%%%%%%%%%%%%%%%%%%%
\subsection{Redshift drift cosmography}

An alternative phenomenological approach to cosmology, first explored by \citet{Visser}, is cosmography, where, with some caveats, physical quantities are expanded as a Taylor series in the cosmological redshift $z$, or some other analogous parameter. A detailed assessment of the cosmological impact and model discriminating power of redshift drift cosmography has been recently reported by \citet{Rocha}, whose methodology we now summarize.

The scale factor can be written as a Taylor expansion up to the sixth order as
\begin{equation} \label{TaylorScale}
\frac{a(t_H)}{a_{\mathrm{0}}} = 1 + t_H - \frac{q_{\mathrm{0}}}{2}(t_H)^2 + \frac{j_{\mathrm{0}}}{3!}(t_H)^3 + \frac{s_{\mathrm{0}}}{4!}(t_H)^4 + \frac{c_{\mathrm{0}}}{5!}(t_H)^5 + \frac{p_{\mathrm{0}}}{6!}(t_H)^6 + O((t_H)^7),
\end{equation}
where we have defined a dimensionless time variable $t_H = H_{\mathrm{0}}(t-t_0)$ and the index $0$ refers to the present day. The cosmographic coefficients are the Hubble constant $H_{\mathrm{0}}$, the deceleration parameter $q_{\mathrm{0}}$, the jerk $j_{\mathrm{0}}$, the snap $s_{\mathrm{0}}$, the crackle $c_{\mathrm{0}}$ and the pop $p_{\mathrm{0}}$, all of them related to the expansion history of the Universe. One of the important results of \citet{Rocha} is that a redshift drift Taylor series in $z$ is vulnerable to biases in the higher order terms, and therefore is reliable  only at low redshifts, but this is not a concern for our present purposes since we will be dealing with redshifts $z<1$.

With these definitions, the Taylor series for the re-scaled Hubble parameter $E(z)$ is given by
\begin{equation} \label{EzCosmography}
\begin{split}
 E(z) & = 1 + (q_{\mathrm{0}} + 1)z + \frac{1}{2}(j_{\mathrm{0}} - q^2_0)z^2 + \frac{1}{3!}(-3j_{\mathrm{0}} -4j_{\mathrm{0}}q_{\mathrm{0}} +3q^3_0 +3q^2_0 -s_{\mathrm{0}})z^3 + \\
 &  + \frac{1}{4!}(c_{\mathrm{0}} -4j^2_0 +25j_{\mathrm{0}}q^2_0+32j_{\mathrm{0}}q_{\mathrm{0}} +12j_{\mathrm{0}} -15q^4_0 -24q^3_0 -12q^2_0 +7q_{\mathrm{0}}s_{\mathrm{0}} +8s_{\mathrm{0}})z^4 +\\
 & + \frac{1}{5!}(-11c_{\mathrm{0}}q_{\mathrm{0}} -15c_{\mathrm{0}} +70j^2_0q_{\mathrm{0}} + 60j^2_0-210j_{\mathrm{0}}q^3_0 -375j_{\mathrm{0}}q^2_0 -240j_{\mathrm{0}}q_{\mathrm{0}} +15j_{\mathrm{0}}s_{\mathrm{0}} -60j_{\mathrm{0}} - \\
 & -p_{\mathrm{0}} +105q^5_0 +225q^4_0 +180q^3_0 -60q^2_0s_{\mathrm{0}} -105q_{\mathrm{0}}s_{\mathrm{0}} +60q^2_0 -60s_{\mathrm{0}})z^5 + O(z^6)\,.
\end{split}
\end{equation}
From this it is straightforward to calculate the cosmographic series for the redshift drift $(1/H_{\mathrm{0}})(\Delta z/\Delta t)$ (cf. Eq. \ref{eq1}) and for the spectroscopic velocity $\Delta v$ (cf. Eq. \ref{specvel}). We refer the reader to \citet{Rocha} for their explicit expressions.

As a way of relating the two approaches, it is instructive to write down the cosmographic coefficients for the CPL parametrization, which has as a particular case the $\Lambda$CDM model when $w_{\mathrm{0}}=-1$ and $w_{\mathrm{a}}=0$. Their generic form is
\begin{equation}\label{q0}
    q_{\mathrm{0}} = \frac{1}{2}(w_{\mathrm{0}}(3-3\Omega_{\mathrm{m}})+1);
\end{equation}
\begin{equation}\label{j0}
    j_{\mathrm{0}} = \frac{1}{2}(w_{\mathrm{a}}(3-3\Omega_{\mathrm{m}})+w_{\mathrm{0}}^2(9-9\Omega_{\mathrm{m}})+w_{\mathrm{0}}(9-9\Omega_{\mathrm{m}})+2);
\end{equation}
\begin{equation}\label{s0}
\begin{split}
    s_{\mathrm{0}} & = \frac{1}{4} (w_{\mathrm{a}}(33\Omega_{\mathrm{m}}-33) + w_{\mathrm{0}}^3(-27\Omega_{\mathrm{m}}^2+108\Omega_{\mathrm{m}}-81)+ w_{\mathrm{0}}^2(-27\Omega_{\mathrm{m}}^2+171\Omega_{\mathrm{m}}-144) + \\
    & + w_{\mathrm{0}}(81\Omega_{\mathrm{m}} +w_{\mathrm{a}}(-9\Omega_{\mathrm{m}}^2+72\Omega_{\mathrm{m}}-63)-81)-14);
\end{split}
\end{equation}
\begin{equation}\label{c0}
\begin{split}
    c_{\mathrm{0}} & = \frac{1}{4} (w_{\mathrm{a}}^2(9\Omega_{\mathrm{m}}^2-72\Omega_{\mathrm{m}}+63) + w_{\mathrm{a}}(213-213\Omega_{\mathrm{m}}) + w_{\mathrm{0}}^4(324\Omega_{\mathrm{m}}^2-810\Omega_{\mathrm{m}}+486)+ \\
    & + w_{\mathrm{0}}^3(648\Omega_{\mathrm{m}}^2-1917\Omega_{\mathrm{m}}+1269) + w_{\mathrm{0}}^2(378\Omega_{\mathrm{m}}^2-1584\Omega_{\mathrm{m}} + w_{\mathrm{a}}(297\Omega_{\mathrm{m}}^2-918\Omega_{\mathrm{m}}+621)+1206)+\\
    & w_{\mathrm{0}}(-489\Omega_{\mathrm{m}} + w_{\mathrm{a}}(189\Omega_{\mathrm{m}}^2-927\Omega_{\mathrm{m}}+738)+489)+70);
\end{split}
\end{equation}
\begin{equation}\label{p0}
\begin{split}
    p_{\mathrm{0}} & = \frac{1}{8} (w_{\mathrm{a}}^2(-459\Omega_{\mathrm{m}}^2+2502\Omega_{\mathrm{m}}-2043)+ w_{\mathrm{a}}(3321\Omega_{\mathrm{m}}-3321)+ w_{\mathrm{0}}^5(972\Omega_{\mathrm{m}}^3-8262\Omega_{\mathrm{m}}^2+14580\Omega_{\mathrm{m}}-7290) \\
    & + w_{\mathrm{0}}^4(1944\Omega_{\mathrm{m}}^3-23409\Omega_{\mathrm{m}}^2+46818\Omega_{\mathrm{m}}-25353)+w_{\mathrm{0}}^3(1134\Omega_{\mathrm{m}}^3-23814\Omega_{\mathrm{m}}^2+57267\Omega_{\mathrm{m}} + w_{\mathrm{a}}(891\Omega_{\mathrm{m}}^3-11745\Omega_{\mathrm{m}}^2 +\\
    & +24057\Omega_{\mathrm{m}}-13203)-34587) +  w_{\mathrm{0}}^2(-9315\Omega_{\mathrm{m}}^2+32328\Omega_{\mathrm{m}}+w_{\mathrm{a}}(567\Omega_{\mathrm{m}}^3-16065\Omega_{\mathrm{m}}^2+41013\Omega_{\mathrm{m}}-25515)-23013)+\\
    & + w_{\mathrm{0}}(7407\Omega_{\mathrm{m}}+w_{\mathrm{a}}^2(27\Omega_{\mathrm{m}}^3-1863\Omega_{\mathrm{m}}^2+5265\Omega_{\mathrm{m}}-3429) + w_{\mathrm{a}}(-5508\Omega_{\mathrm{m}}^2+21645\Omega_{\mathrm{m}}-16137)-7407)-910)\,.
\end{split}
\end{equation}

Two remarks are in order at this point. The first is that an observational measurement finding $j_{\mathrm{0}}\neq1$, no matter at what redshift, would immediately disprove the flat $\Lambda$CDM model. The second emerges when one inverts the above expressions in order to express the CPL model parameters as a function of the cosmographic ones. One finds
\begin{equation}
\Omega_{m} = \frac{8j_{\mathrm{0}}q_{\mathrm{0}} + 7j_{\mathrm{0}} - 16q_{\mathrm{0}}^2 + 2q_{\mathrm{0}} + 2s_{\mathrm{0}} - 1 + (1-2q_{\mathrm{0}})\sqrt{9j_{\mathrm{0}}^2 + 8j_{\mathrm{0}}q_{\mathrm{0}}^2 - 12j_{\mathrm{0}}q_{\mathrm{0}} - 14j_{\mathrm{0}} + 32q_{\mathrm{0}}^2 + 8q_{\mathrm{0}}s_{\mathrm{0}} - 4s_{\mathrm{0}} + 1}}{2j_{\mathrm{0}}q_{\mathrm{0}} + 10j_{\mathrm{0}} - 8q_{\mathrm{0}} + 2s_{\mathrm{0}}}
\end{equation}
\begin{equation}
w_{\mathrm{0}} = \frac{1-2q_{\mathrm{0}}}{3(\Omega_{m}-1)}
\end{equation}
\begin{equation}
w_{\mathrm{a}} = \frac{(-2j_{\mathrm{0}}+6q_{\mathrm{0}}-1)\Omega_{m} + 2j_{\mathrm{0}}-4q_{\mathrm{0}}^2-2q_{\mathrm{0}}}{3(\Omega_{m}-1)^2}\,.
\end{equation}
This shows that for the flat CPL model to be fully specified one needs to know only the cosmographic parameters $q_{\mathrm{0}}$, $j_{\mathrm{0}}$ and $s_{\mathrm{0}}$. Moreover, if one assumes that the matter density is known, $q_{\mathrm{0}}$, $j_{\mathrm{0}}$ are sufficient to determine the dark energy equation of state. In contrast, the parameters $c_{\mathrm{0}}$ and $p_{\mathrm{0}}$ are not needed, although they could (if they can be precisely and accurately determined) provide consistency tests for such a model.

%%%%%%%%%%%%%%%%%%%%%%%%%%%%%%%%%%%%%%%%%%%%%%%%%%%%%%%%%%%%%%%%%%%%%%%%%%%%%%%%%
\subsection{Higher-order derivatives}
\label{hods}

In principle, the SKAO enables the further possibility  of measurements of time or redshift derivatives of the redshift drift. Other than being obtained by direct means (for which a specific implementation is lacking) this drift of the drift can be obtained numerically from a set of measurements of the redshift drift at different redshifts, as has been discussed in \citet{Second}.

In this context it is more convenient to use a slightly different notation
\begin{equation}
\label{eq:Z1}
    Z_{\mathrm{1}}(z) = \frac{1}{H_{\mathrm{0}}}\frac{dz}{dt_0} = 1 + z - E(z)\,,
\end{equation}
with $Z_{\mathrm{1}}$ denoting the first dimensionless time derivative of the redshift, which is related to the spectroscopic velocity via
\begin{equation}
    \Delta v (z)= cH_{\mathrm{0}}\Delta t\,\frac{Z_{\mathrm{1}}(z)}{1+z}\,.
\end{equation}
A practical advantage of this choice is that one can then similarly define the second dimensionless time derivative; this has the form
\be
Z_{\mathrm{2}}(z)=\frac{1}{H_{\mathrm{0}}^2}\frac{d^2z}{dt_0^2}=\frac{1+q(z)}{1+z}E^2(z)-E(z)-q_{\mathrm{0}}(1+z) \,,
\ee
or equivalently
\be\label{eq:z2}
Z_{\mathrm{2}}(z)=\frac{1+q(z)}{1+z}Z_{\mathrm{1}}^2(z)-\left(1+2q(z)\right)Z_{\mathrm{1}}+\left(q(z)-q_{\mathrm{0}}\right)(1+z)\,,
\ee
where we also used the definition of the generic (redshift-dependent) deceleration parameter
\be
q(z)=-\frac{a{\ddot a}}{{\dot a}^2}=-\frac{\ddot a}{aH^{2}(a)}=-1+\frac{1}{2}(1+z)\frac{[E^2(z)]'}{E^2(z)}\,,
\ee
with the further convention that $'\equiv d/dz$. Note that with these definitions the Hubble constant is not among the free parameters. It is then straightforward to show that in the low redshift limit the first derivative can be written
\be\label{eq:lowz1}
Z_{\mathrm{1}}=-q_{\mathrm{0}}z+O(z^2) \,,
\ee
while for the second derivative we have
\be\label{eq:lowz2}
Z_{\mathrm{2}}=j_{\mathrm{0}}z+O(z^2)\,.
\ee
In other words, measurements of $Z_{\mathrm{1}}$ and $Z_{\mathrm{2}}$ directly yield, at least in principle, the present-day values of the deceleration and jerk parameters. Note that previous studies, not specific to the SKAO, do suggest that the redshift drift is an optimal way to measure the deceleration parameter \citep{Neben}, because it allows a direct measurement of $q_{\mathrm{0}}$ with both accuracy and precision, which is not possible using traditional distance indicators. Specifically, at low redshifts a Taylor expansion is accurate but there is little leverage on $q_{\mathrm{0}}$, and additionally one faces the limitations caused by significant peculiar velocities. Conversely, at higher redshifts the peculiar velocities are not an important consideration but the Taylor expansion approximation will be less accurate and introduces a bias, and one must also take into account evolutionary effects, which can significantly impact the error budget.

In practice there is currently no feasible way of directly measuring $Z_{\mathrm{2}}$. Nevertheless, there is a quantity closely related to $Z_{\mathrm{2}}$ which can be obtained through the numerical redshift derivative of $Z_{\mathrm{1}}$, namely
\be
\frac{dZ_{\mathrm{1}}(z)}{dz} = 1 - E(z)'= -q(z)+Z_{\mathrm{1}}(z) \frac{1+q(z)}{1+z} = 1+\frac{1+q(z)}{1+z} [Z_{\mathrm{1}}-(1+z)]\,, \label{eq:z1der}
\ee
where the definition of the deceleration parameter has been used. In the low-redshift limit $Z_{\mathrm{2}}$ and $dZ_{\mathrm{1}}/dz$, although different, contain the same cosmographic information. Specifically, to linear order in redshift $dZ_{\mathrm{1}}/dz$ becomes
\be\label{eq:lowzD}
\frac{dZ_{\mathrm{1}}(t_0,z)}{dz} = -q_{\mathrm{0}}+(q_{\mathrm{0}}^2-j_{\mathrm{0}})z+O(z^2)\,,
\ee
which again depends only on the deceleration and jerk parameters. We emphasize that this expression is fully generic (other than the assumption of a metric theory of gravity), and therefore the constraints coming from its measurement would be fully model independent. If one has $N$ measurements of $\Delta v=(cH_{\mathrm{0}}\Delta t)Z_{\mathrm{1}}/(1+z)$ at some redshifts $z_i$, one can obtain $N-1$ correlated measurements of $dZ_{\mathrm{1}}/dz$ as numerical derivatives
\be\label{eq:dz1dz}
D(\bar{z})\equiv\frac{dZ_{\mathrm{1}}}{dz}(\bar{z}) = \frac{Z_{\mathrm{1}}(z_{i+1})-Z_{\mathrm{1}}(z_i)}{z_{i+1}-z_i}
\ee
with $\bar{z}=(z_{i+1}-z_i)/2$.\\
The errors on these measurements can be obtained by standard error propagation
\be\label{eq:Derr}
\frac{\sigma^2_D}{D^2}=\frac{\sigma^2_{Z_{\mathrm{1}}(z_{i+1})}+\sigma^2_{Z_{\mathrm{1}}(z_{i})}}{(Z_{\mathrm{1}}(z_{i+1})-Z_{\mathrm{1}}(z_i))^2} + \frac{\sigma^2_{z_{i+1}}+\sigma^2_{z_{i}}}{(z_{i+1}-z_i)^2}
\ee
with $\sigma_{Z_{\mathrm{1}}(z_{i})}$ correspondingly obtained by propagating the error on $\Delta v$; in what follows we will generally ignore the second term, which according to the results of \citet{Second}, is expected to be entirely subdominant. Figure \ref{fig1} illustrates the redshift dependence of $Z_{\mathrm{1}}$, $Z_{\mathrm{2}}$ and $D$ for three flat $\Lambda$CDM models with different matter densities; note that all three quantities are independent of the value of the Hubble constant. It is also worthy of note that $Z_{\mathrm{1}}$ vanishes at redshift $z=0$ (by definition, as does $Z_{\mathrm{2}}$) and again at $z\sim2$, being positive below the latter redshift and negative above it. This is to be contrasted with $D$, which has a value $D_0=-q_{\mathrm{0}}$ at redshift zero, vanishes around $z\sim1$ and becomes negative at higher redshifts. One consequence of this behaviour is that obtaining $D(z)$ at redshifts $z\sim1$ though numerical derivatives as discussed above will imply large uncertainties, as will be discussed below.

\begin{figure*}
\begin{center}
\includegraphics[width=0.32\columnwidth,keepaspectratio]{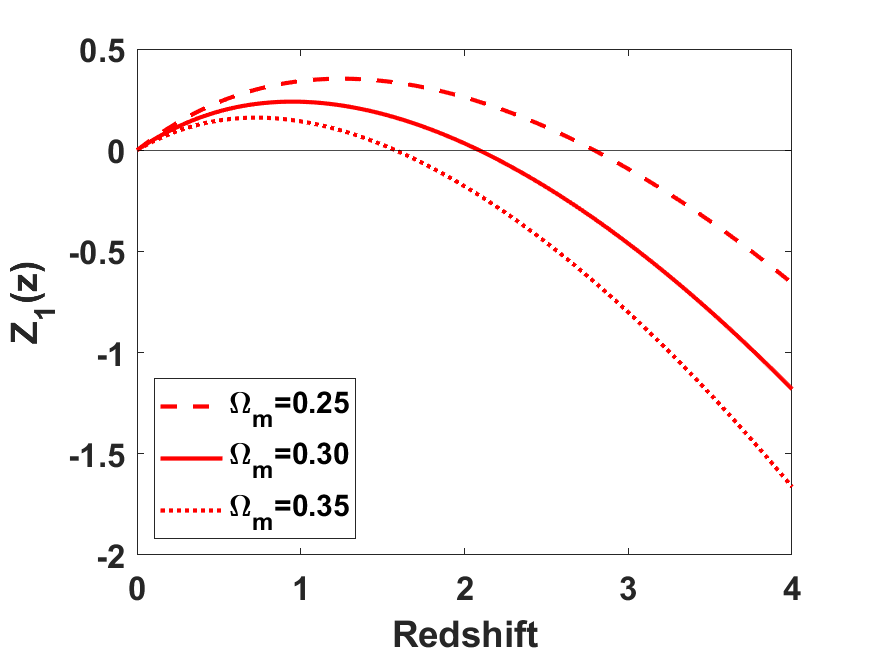}
\includegraphics[width=0.32\columnwidth,keepaspectratio]{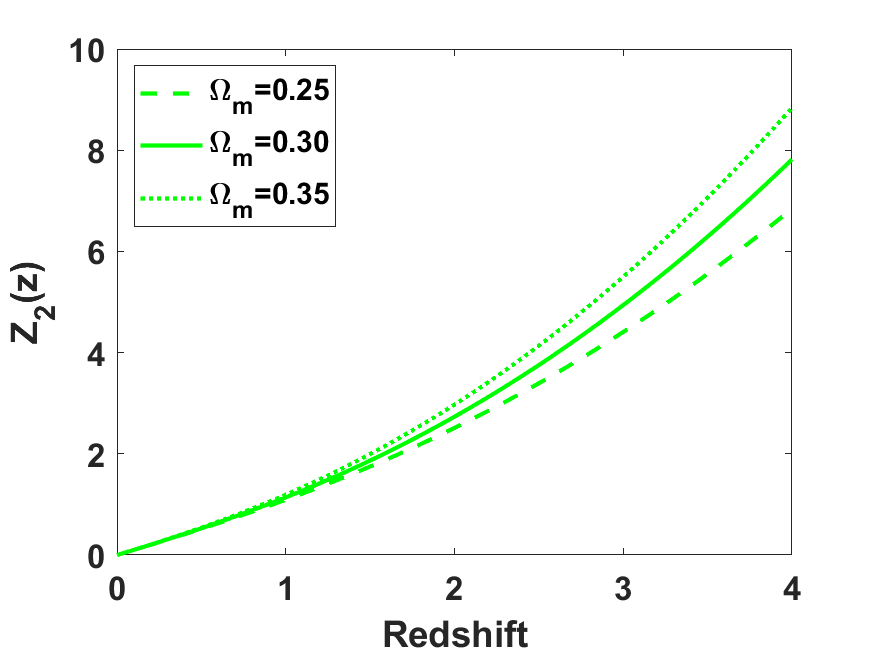}
\includegraphics[width=0.32\columnwidth,keepaspectratio]{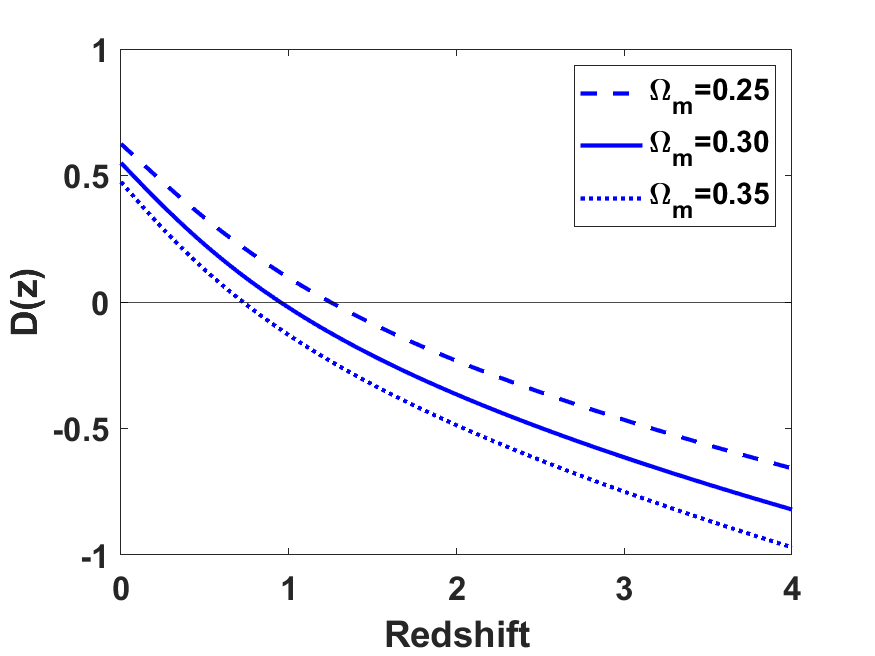}
\end{center}
\caption{Redshift dependence of $Z_{\mathrm{1}}(z)$, $Z_{\mathrm{2}}(z)$ and $D(z)$, defined in the main text, for three flat $\Lambda$CDM models with different matter densities. For convenience, the dotted horizontal lines identify the vanishing values of the plotted quantities.}
\label{fig1}
 \end{figure*}

%%%%%%%%%%%%%%%%%%%%%%%%%%%%%%%%%%%%%%%%%%%%%%%%%%%%%%%%%%%%%%%%%%%%%%%%%%%%%%%%%
%%%%%%%%%%%%%%%%%%%%%%%%%%%%%%%%%%%%%%%%%%%%%%%%%%%%%%%%%%%%%%%%%%%%%%%%%%%%%%%%%

\section{Robustness of the Fisher Matrix approach}

We start by exploring the robustness of the Fisher Matrix based forecasts, by comparing its results for the cosmographic series with those obtained using an MCMC approach in \citet{Rocha}. Following the analysis of \citet{Klockner} we assume $10$ SKAO redshift drift measurements, equally spaced between redshifts $z = 0.1$ and $z = 1.0$ and with the associated spectroscopic velocity uncertainties equally spaced between $1\%$ and $10\%$ respectively. We assume the aforementioned flat $\Lambda$CDM fiducial model, which corresponds to the cosmographic series coefficients $H_{\mathrm{0}} = 70$ km/s/Mpc, $q_{\mathrm{0}} = -0.55$, $j_{\mathrm{0}} = 1$, $s_{\mathrm{0}} = -0.35$, $c_{\mathrm{0}} = 3.115$, $p_{\mathrm{0}} = -10.88675$. Moreover, we assume the following priors for each of the coefficients: $\sigma_{H_{\mathrm{0}}} = 10$ km/s/Mpc, $\sigma_{q_{\mathrm{0}}} = 1$, $\sigma_{j_{\mathrm{0}}} = 10$, $\sigma_{s_{\mathrm{0}}} = 30$, $\sigma_{c_{\mathrm{0}}} = 50$ and $\sigma_{p_{\mathrm{0}}} = 100$. These assumptions are akin to those in \citet{Rocha} and therefore enable a comparison of the forecast constraints obtained in the two approaches.

%%%%%%%%%%%%%%%%%%%%%%%%%%%%%%%%%%%%%%%%%
\begin{table}
    \centering
    \caption{One-sigma uncertainties derived from the Fisher Matrix analysis of the cosmographic series with four and six parameters for a flat $\Lambda$CDM fiducial model, compared to the corresponding MCMC results obtained in \citet{Rocha}.}
    \label{tab1}
\begin{tabular}{llrrrr}
\toprule
\multicolumn{1}{|l|}{\textbf{Cosmographic}} & \multicolumn{1}{|l|}{\textbf{Fiducial}} & \multicolumn{1}{|r|}{\textbf{Prior}} & \multicolumn{2}{|c|}{\textbf{FMA uncertainties}} & \multicolumn{1}{|c|}{\textbf{MCMC}} \\
\textbf{Parameter} & \textbf{Value} & \textbf{Uncertainty} & \textbf{6 parameters} &  \textbf{4 parameters} &  \textbf{6 parameters} \\
\toprule
$\boldsymbol{H_{\mathrm{0}}}$ $[km/s/Mpc]$ &    70  & 10    &  9.97 &        9.97  &  67.21\tiny{\enskip \Vectorstack{+7.63 -5.23}}\\
$\boldsymbol{q_{\mathrm{0}}}$ &    -0.55  & 1  &  0.08 &         0.08 &  -0.58\tiny{\enskip \Vectorstack{+0.06 -0.05}} \\
$\boldsymbol{j_{\mathrm{0}}}$ &    1   & 10    &  0.67 &         0.23 &   1.22\tiny{\enskip \Vectorstack{+0.59 -0.52}} \\
$\boldsymbol{s_{\mathrm{0}}}$ &    -0.35   & 30 &  6.42 &         0.38 &   1.53\tiny{\enskip \Vectorstack{+6.49 -5.18}} \\
$\boldsymbol{c_{\mathrm{0}}}$ &    3.115  & 50   & 16.21 &             - &   9.45\tiny{\enskip \Vectorstack{+19.96 -14.68}} \\
$\boldsymbol{p_{\mathrm{0}}}$ &    -10.88675 & 100 & 70.04 &             - &   1.04\tiny{\enskip \Vectorstack{+54.41 -47.70}} \\
\bottomrule
\end{tabular}
\end{table}
%%%%%%%%%%%%%%%%%%%%%%%%%%%%%%%%%%%%%%%%%%%%%%

Table \ref{tab1} shows the results of this comparison, listing the one-sigma uncertainties by our Fisher Matrix Analysis (FMA) and the corresponding MCMC results. Overall, we find reasonable agreement between the two approaches, with the Fisher Matrix slightly overestimating the uncertainties. In absolute terms the difference is clearest in the case of the Hubble constant, in which case there is almost no improvement on the prior---a result of the fact that in a mathematical sense $H_{\mathrm{0}}$ is purely a multiplicative factor for the redshift drift.

The table also shows the impact of truncating the series at fourth rather than sixth order (a point to which we will return later in this work). In this case the constraints on the Hubble and deceleration parameters are unaffected, but those on the jerk and the snap would be significantly improved, by factors of about 3 and 17 respectively. The reason for these differences is the more complex degeneracies between the higher order parameters in the cosmographic series. Figure \ref{fig2} illustrates some of these degeneracies in relevant 2D planes.

\begin{figure*}
\begin{center}
\includegraphics[width=0.32\columnwidth,keepaspectratio]{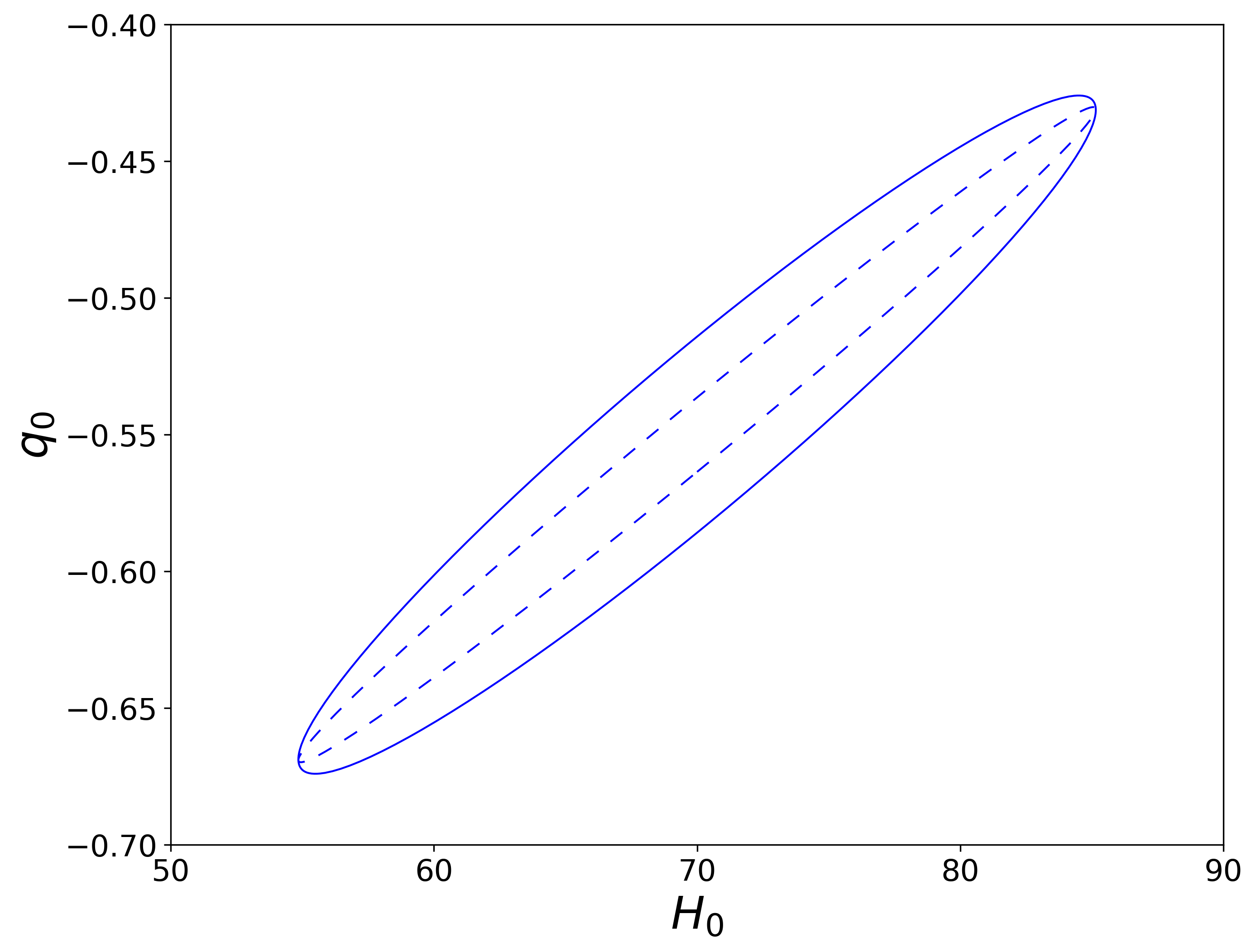}
\includegraphics[width=0.32\columnwidth,keepaspectratio]{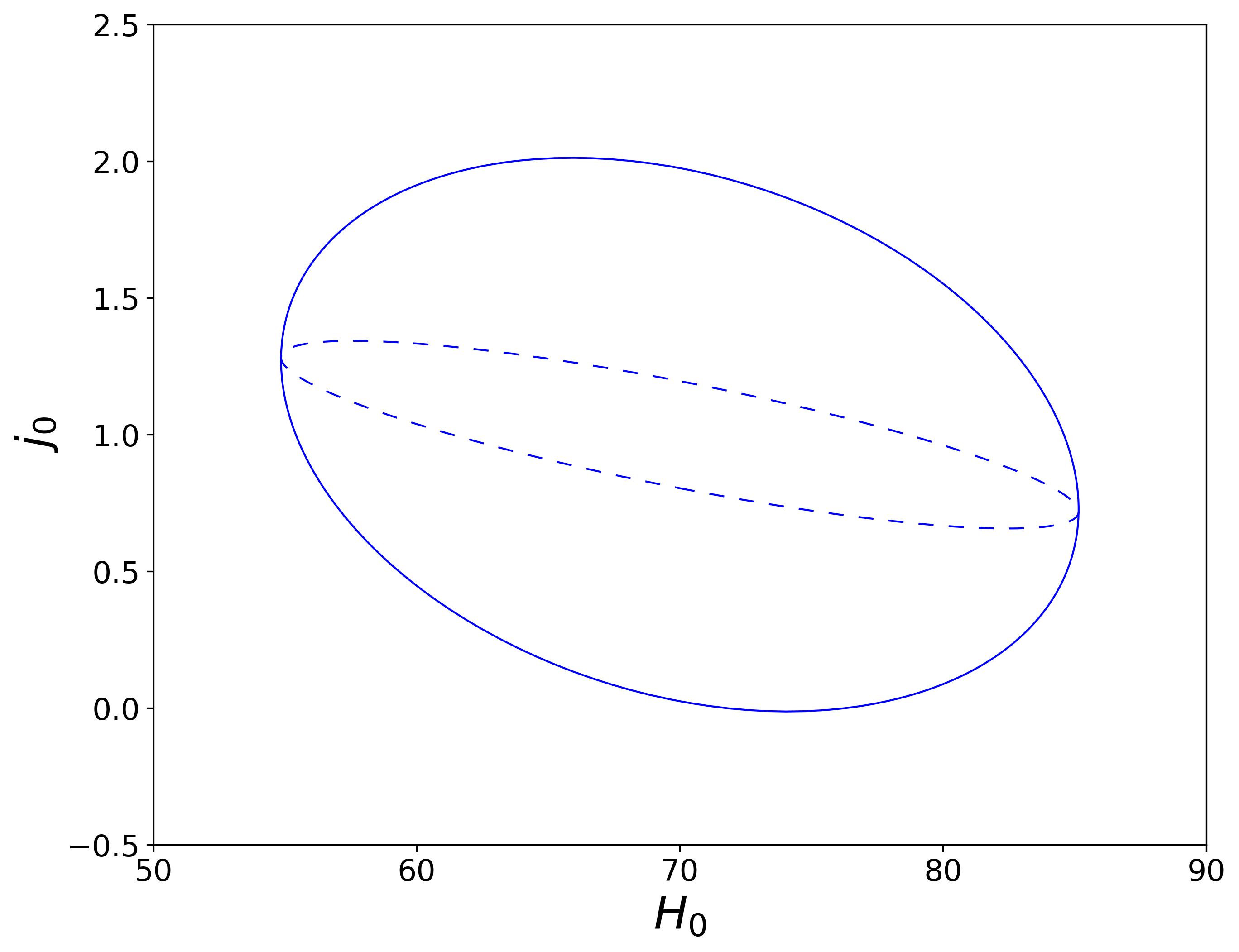}
\includegraphics[width=0.32\columnwidth,keepaspectratio]{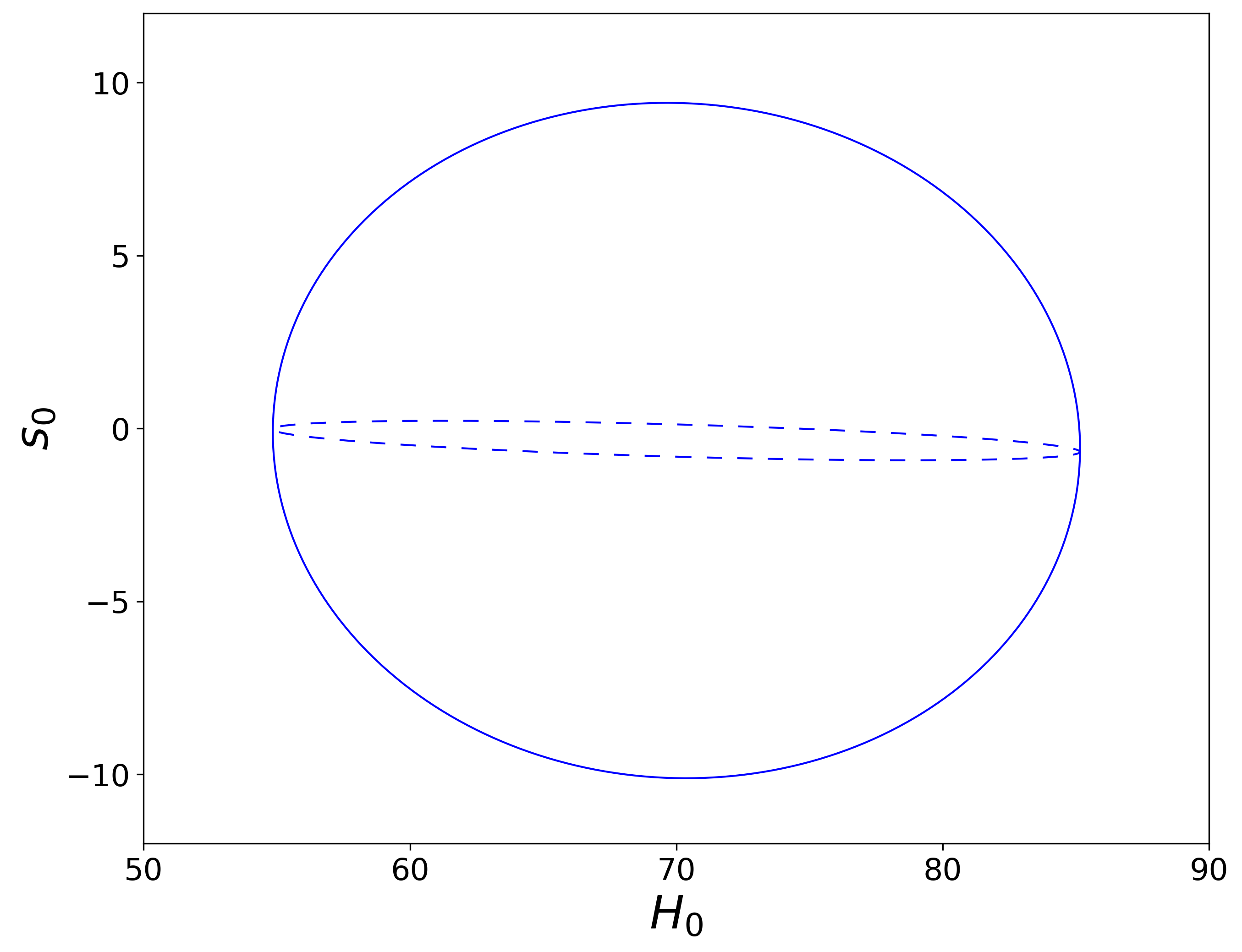}
\includegraphics[width=0.32\columnwidth,keepaspectratio]{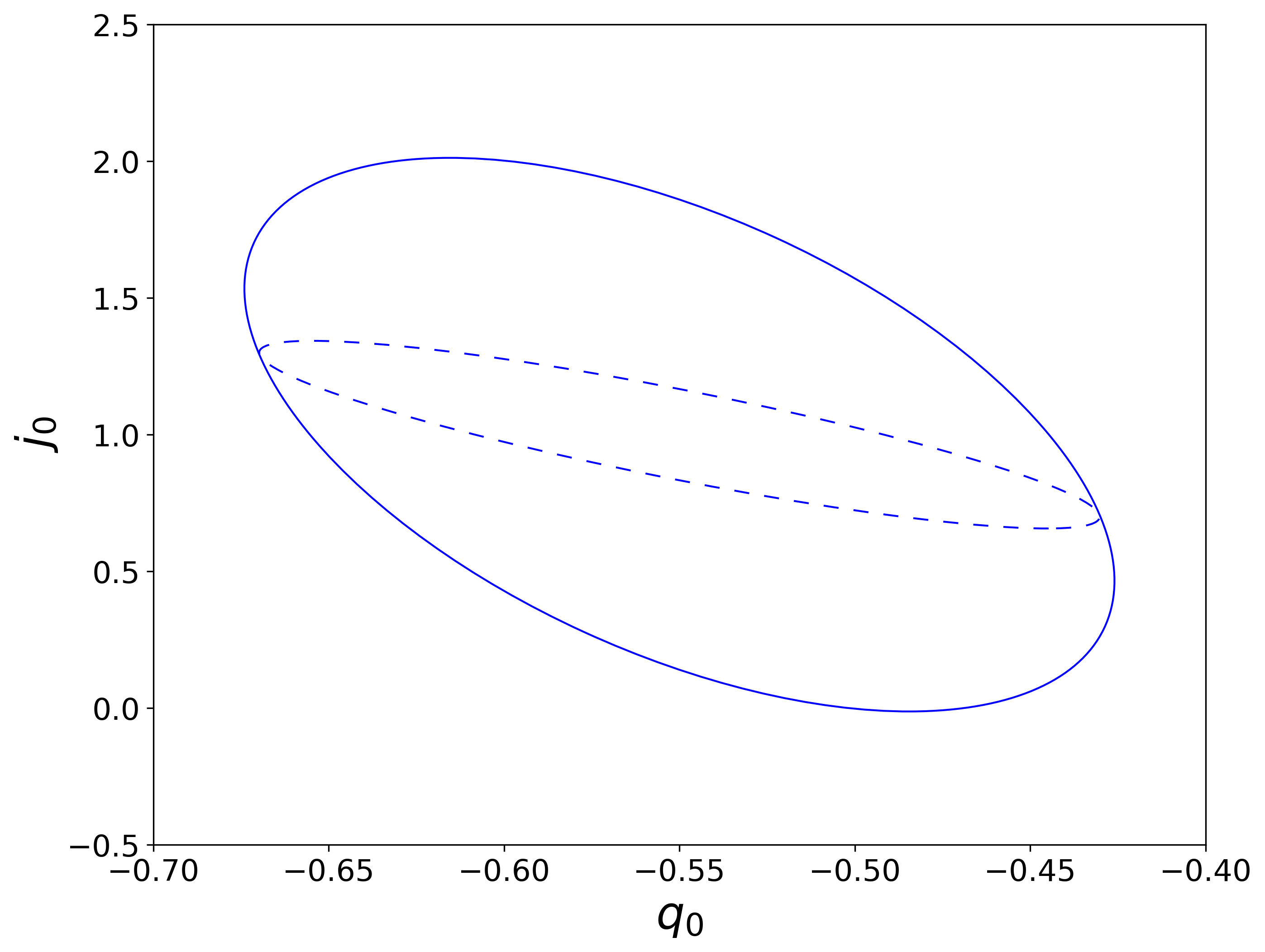}
\includegraphics[width=0.32\columnwidth,keepaspectratio]{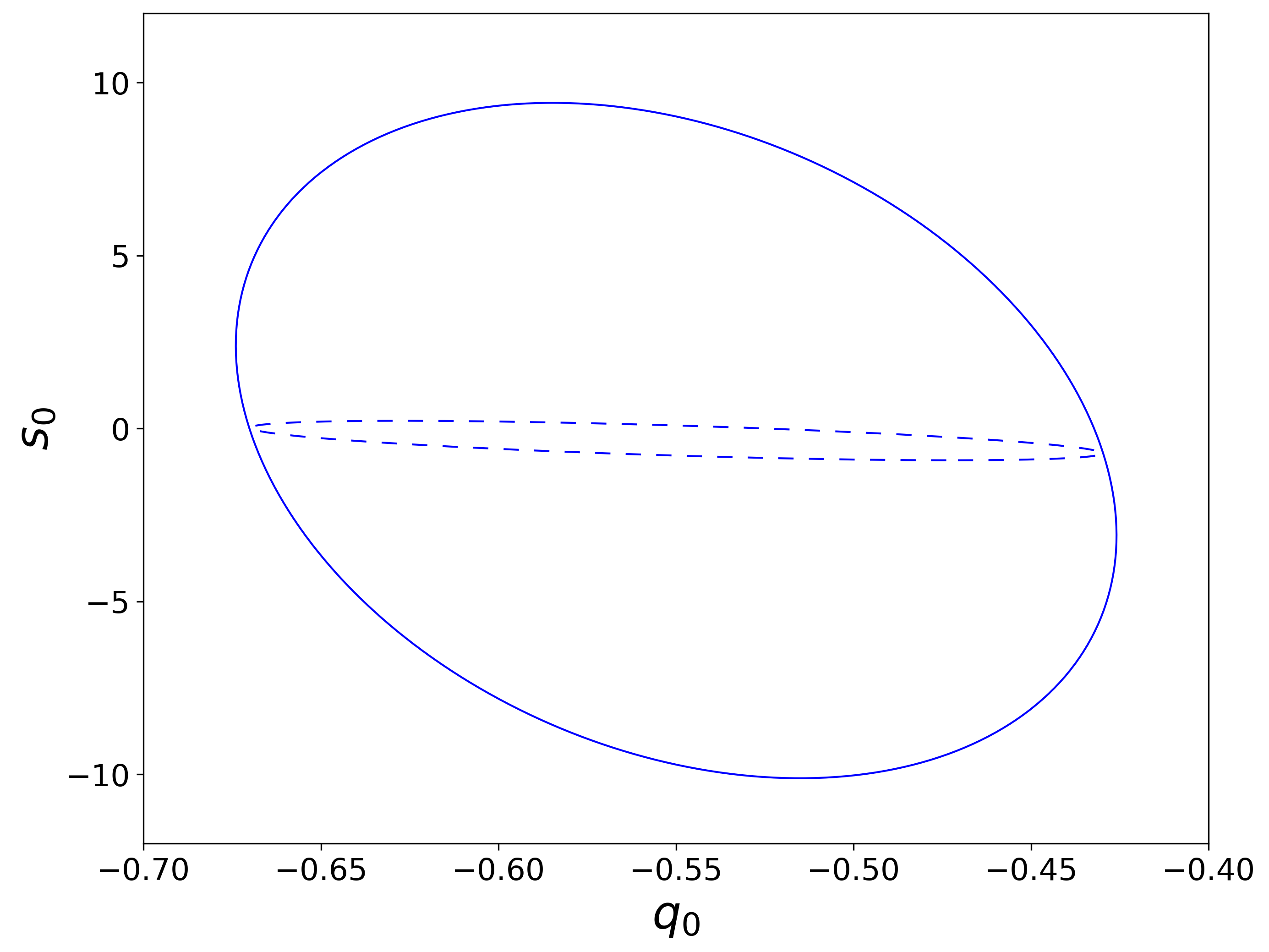}
\includegraphics[width=0.32\columnwidth,keepaspectratio]{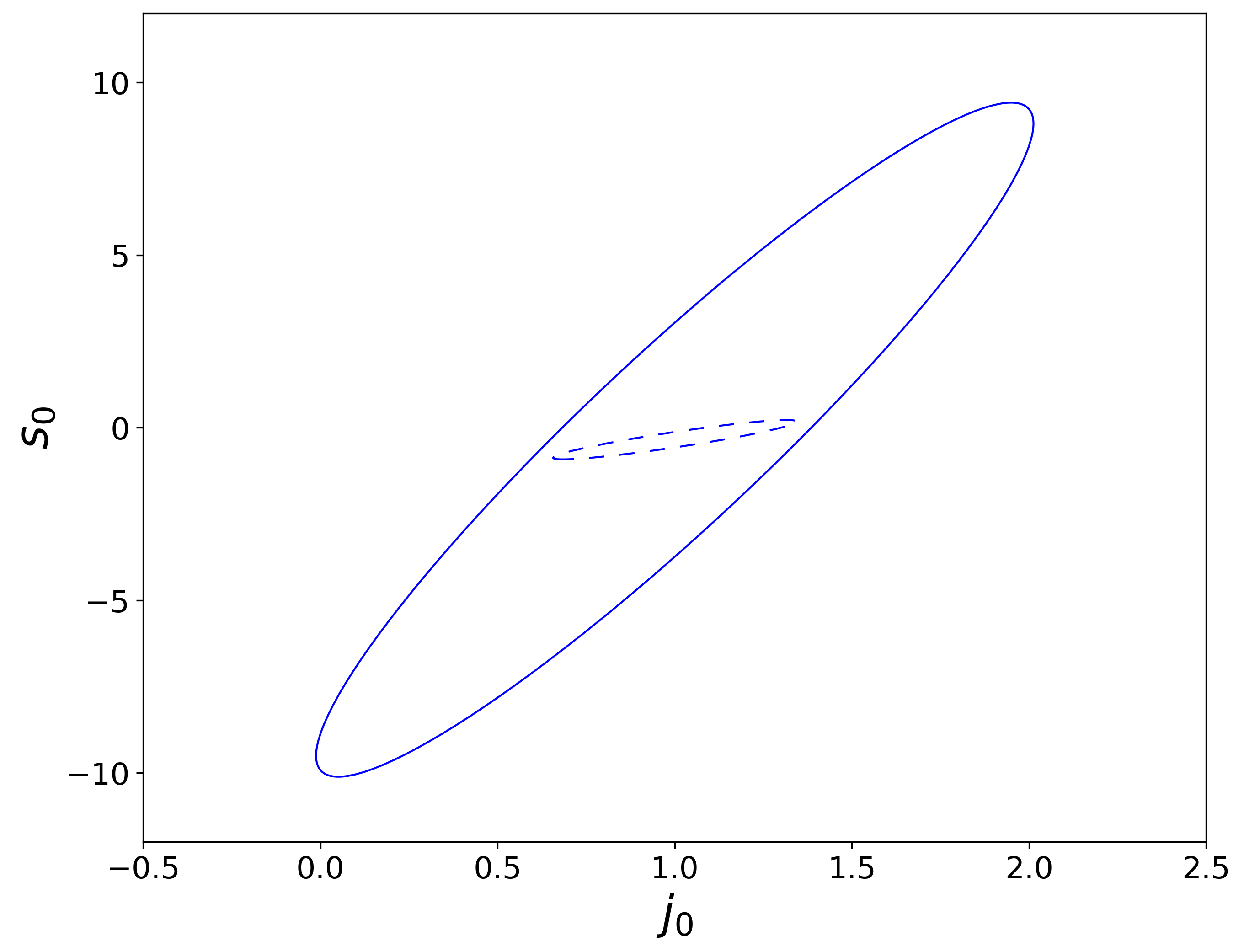}
\includegraphics[width=0.32\columnwidth,keepaspectratio]{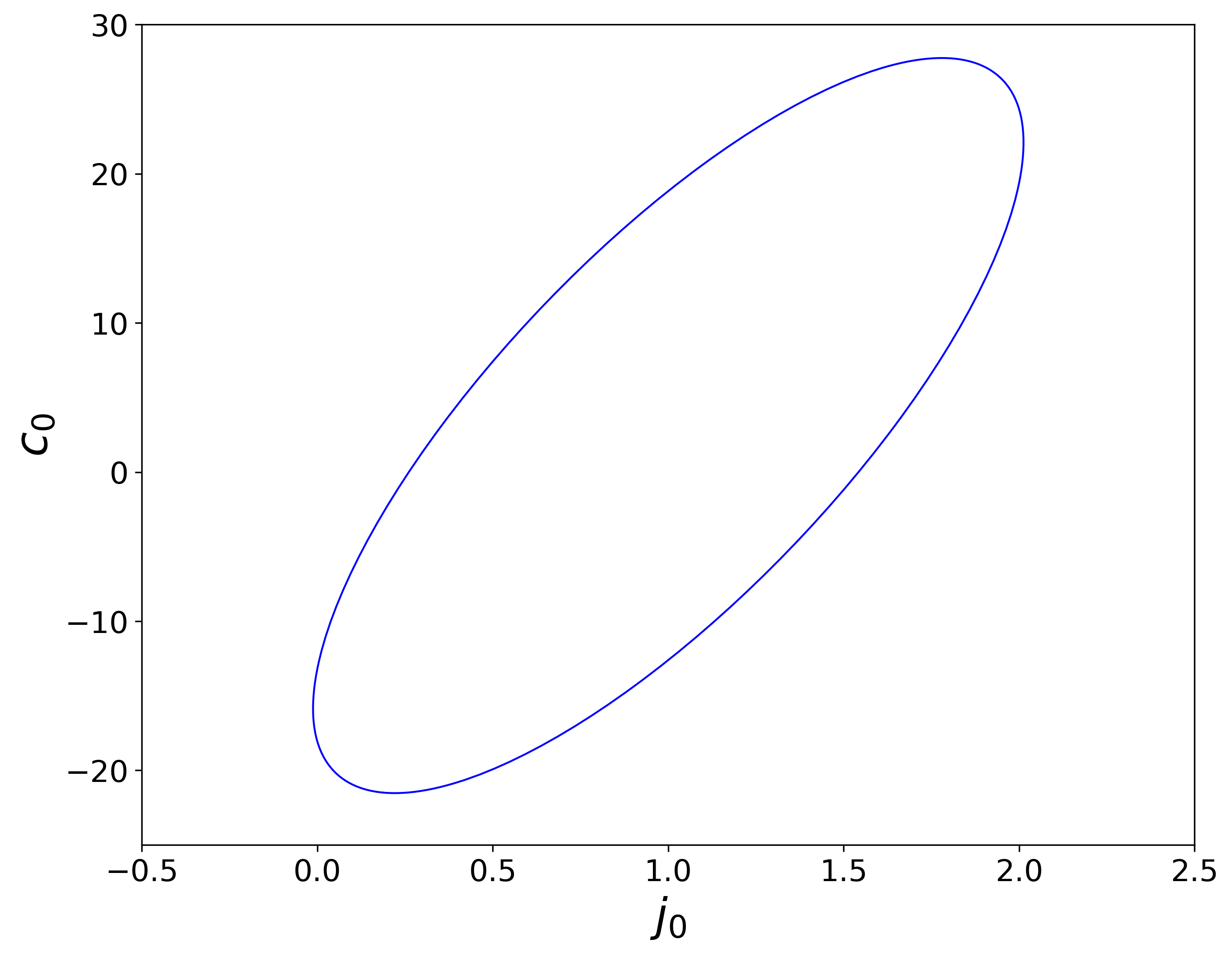}
\includegraphics[width=0.32\columnwidth,keepaspectratio]{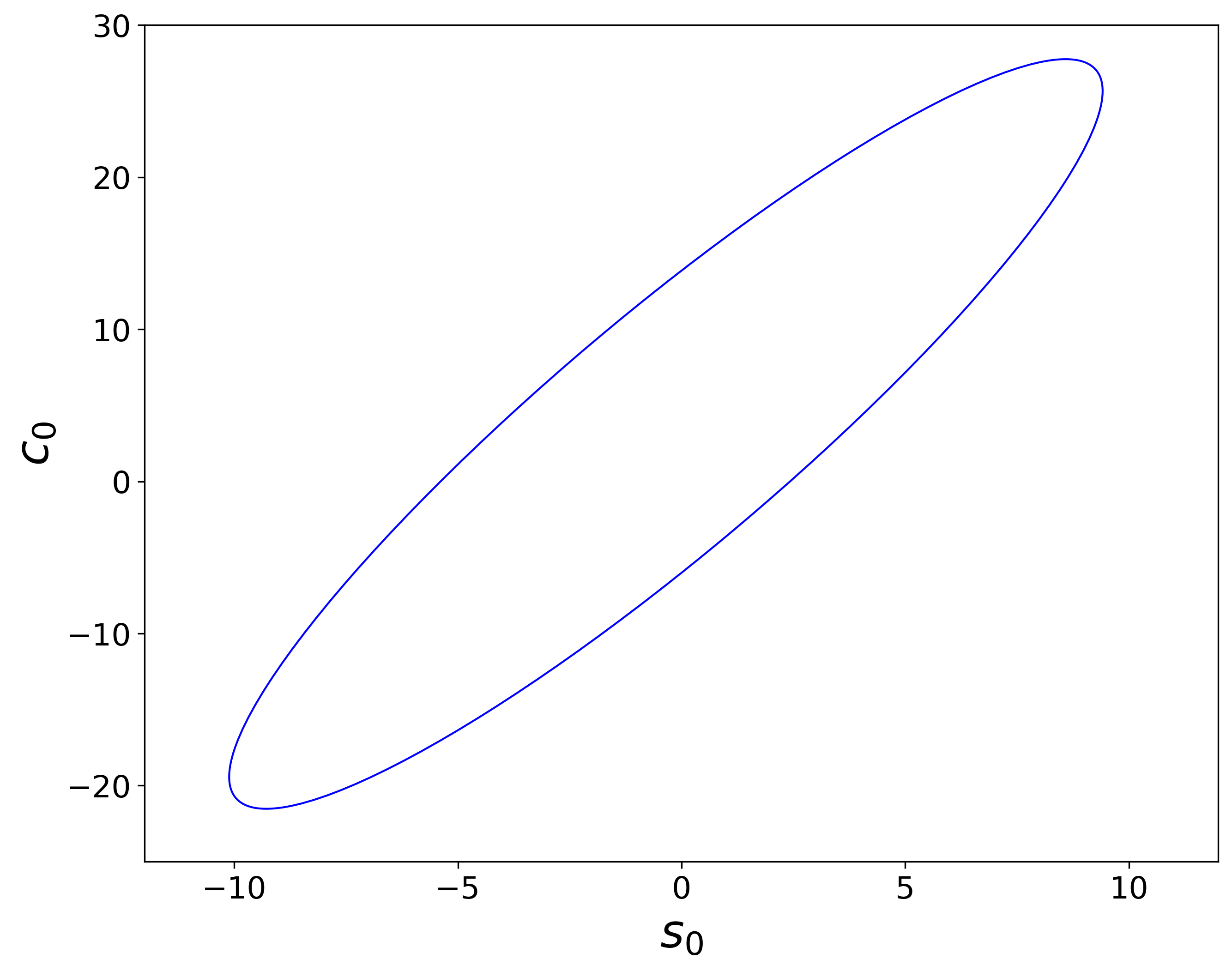}
\includegraphics[width=0.32\columnwidth,keepaspectratio]{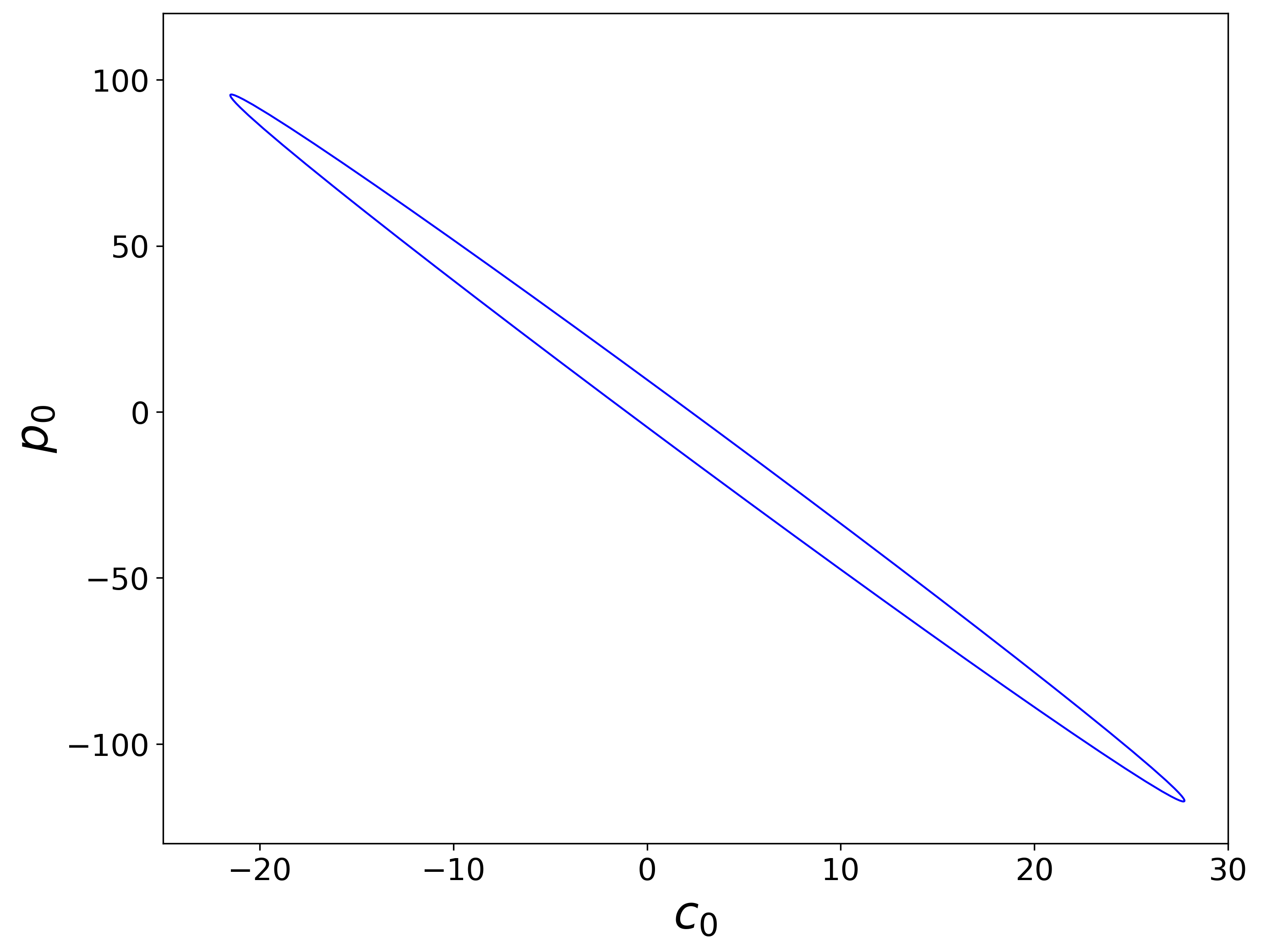}
\end{center}
\caption{One-sigma constraints in relevant 2D planes for the cosmographic series analysis of the $\Lambda$CDM model described in the text. Solid and dashed lines are for the cosmographic series with six parameters and with four parameters respectively. Note the different degeneracy directions in the two cases.}
\label{fig2}
 \end{figure*}

%%%%%%%%%%%%%%%%%%%%%%%%%%%%%%%%%%%%%%%%%%%%%%%%%%%%%%%%%%%%%%%%%%%%%%%%%%%%%%%%%
%%%%%%%%%%%%%%%%%%%%%%%%%%%%%%%%%%%%%%%%%%%%%%%%%%%%%%%%%%%%%%%%%%%%%%%%%%%%%%%%%

\section{Impact of the choice of priors}

Having shown, in the previous section, that Fisher Matrix based forecasts yield results in good agreement with those of an MCMC approach, in this section we rely on the favourable computational speed of the former to carry out a thorough exploration of the impact of the choice of priors on the resulting forecasts. We assume as baseline the fiducial model and choice of priors already mentioned in the previous section (and also listed in Table \ref{tab1}) and study the impact of separately widening or narrowing each of these priors by a factor of ten, while keeping the other priors unchanged. We will refer to the cases where the prior uncertainties are ten times larger and smaller as the pessimistic and optimistic cases, respectively. Table \ref{tab2} summarizes the results of this analysis.

%%%%%%%%%%%%%%%%%%%%%%%%%%%%%%%%%%%%%%%%%
\begin{table}
    \centering
    \caption{Effect of choice of priors on the one-sigma uncertainties derived from the Fisher Matrix analysis for the cosmographic series parameters up to sixth order. The first two lines show the baseline choice of priors and the corresponding result for the baseline fiducial model; both of these are described in the text and also listed in Table \ref{tab1}. The subsequent pairs of lines show, for the same fiducial model, the results for the cases where the prior uncertainty on each parameter is increased or decreased by a factor of 10 with respect to its baseline value (with other priors being unchanged). These are dubbed pessimistic and optimistic priors, respectively.}
    \label{tab2}
\begin{tabular}{lcccccc}
\toprule
\textbf{Prior Choice} & $\boldsymbol{\sigma_{H_{\mathrm{0}}}}$ $[km/s/Mpc]$ & $\boldsymbol{\sigma_{q_{\mathrm{0}}}}$ & $\boldsymbol{\sigma_{j_{\mathrm{0}}}}$ & $\boldsymbol{\sigma_{s_{\mathrm{0}}}}$ & $\boldsymbol{\sigma_{c_{\mathrm{0}}}}$ & $\boldsymbol{\sigma_{p_{\mathrm{0}}}}$ \\
\toprule
Baseline Prior (cf. Table \ref{tab1}) & 10 & 1 &  10 & 30 & 50 & 100\\
Baseline Constraint (cf. Table \ref{tab1}) & 9.97 & 0.08 &  0.67 & 6.42 & 16.21 & 70.04\\
\toprule
Pessimistic $\sigma_{H_{\mathrm{0}}}=100$ km/s/Mpc & 77.75 & 0.61 &  1.54 & 6.53 & 16.24 & 70.04\\
Optimistic $\sigma_{H_{\mathrm{0}}}=1$ km/s/Mpc & 1.00 & 0.02 &  0.64 & 6.42 & 16.21 & 70.04\\
\toprule
Pessimistic $\sigma_{q_{\mathrm{0}}}=10$ & 10.00 & 0.08 &  0.67 & 6.42 & 16.21 & 70.05\\
Optimistic $\sigma_{q_{\mathrm{0}}}=0.1$ & 7.95 & 0.06 &  0.63 & 6.32 & 16.07 & 69.54\\
\toprule
Pessimistic $\sigma_{j_{\mathrm{0}}}=100$ & 9.97 & 0.08 &  0.67 & 6.44 & 16.23 & 70.12\\
Optimistic $\sigma_{j_{\mathrm{0}}}=1$ & 9.86 & 0.08 &  0.56 & 5.49 & 14.67 & 63.85\\
\toprule
Pessimistic $\sigma_{s_{\mathrm{0}}}=300$ & 9.97 & 0.08 &  0.68 & 6.57 & 16.53 & 71.38\\
Optimistic $\sigma_{s_{\mathrm{0}}}=3$ & 9.97 & 0.08 &  0.35 & 2.73 & 9.08 & 40.47\\
\toprule
Pessimistic $\sigma_{c_{\mathrm{0}}}=500$ & 9.97 & 0.08 &  0.69 & 6.73 & 17.13 & 73.98\\
Optimistic $\sigma_{c_{\mathrm{0}}}=5$ & 9.97 & 0.08 &  0.45 & 3.12 & 4.80 & 21.23\\
\toprule
Pessimistic $\sigma_{p_{\mathrm{0}}}=1000$ & 9.97 & 0.08 &  0.82 & 8.54 & 22.58 & 97.66\\
Optimistic $\sigma_{p_{\mathrm{0}}}=10$ & 9.97 & 0.08 &  0.45 & 2.90 & 2.55 & 9.95\\
\bottomrule
\end{tabular}
\end{table}
%%%%%%%%%%%%%%%%%%%%%%%%%%%%%%%%%%%%%%%%%%%%%%

The first salient result is that the Hubble constant is effectively determined by the prior. In other words, there is effectively no improvement on the uncertainty given by the prior, unless one chooses a totally uninformative prior, as in our pessimistic case. The only exception to this behaviour occurs if one uses an optimistic prior for the deceleration parameter, in which case the Hubble constant prior is improved by about 20 percent. Overall, this behaviour is not at all surprising, for precisely the same reason stated in the previous section: the Hubble constant is merely a multiplicative factor in the redshift drift.

The constraint on the deceleration parameter is affected only by the prior on the Hubble constant (except for a minor effect from an optimistic choice of the jerk prior). Remarkably, the posterior constraint on $q_{\mathrm{0}}$ is rather insensitive to the choice of its own prior, and also (with the caveat mentioned in the previous sentence) to the choices of priors on the higher-order parameters. This supports and complements the results of \citet{Neben} identifying the redshift drift as an optimal probe to precisely and accurately constrain the deceleration parameter. On the other hand, behaviour is to be contrasted with the behaviour for the jerk parameter, $j_{\mathrm{0}}$. Its derived uncertainty is changed by a pessimistic prior on $H_{\mathrm{0}}$, relatively unaffected by its own prior and the one on $q_{\mathrm{0}}$, but it is improved by optimistic priors on the higher order parameters (i.e., the snap, crackle, and pop). Finally, the snap, crackle and pop display qualitatively similar behaviours: they are rather insensitive to the priors on the three lower-order parameters ($H_{\mathrm{0}}$, $q_{\mathrm{0}}$ and $j_{\mathrm{0}}$) but sensitive to the priors on the three higher-order parameters---including their own priors, and especially so if the optimistic priors are used.

In general, this analysis suggests that our baseline choice of priors is reasonable, at least to the extent that the constraints derived from this choice are closer to those obtained for the pessimistic cases than they are to the optimistic cases. Moreover, the results for the Hubble constant provide a further pragmatic reason for removing $H_{\mathrm{0}}$ from the free parameters, and working with $Z_{\mathrm{1}}$ and $D$, and we do in the following sections.

%%%%%%%%%%%%%%%%%%%%%%%%%%%%%%%%%%%%%%%%%%%%%%%%%%%%%%%%%%%%%%%%%%%%%%%%%%%%%%%%%
%%%%%%%%%%%%%%%%%%%%%%%%%%%%%%%%%%%%%%%%%%%%%%%%%%%%%%%%%%%%%%%%%%%%%%%%%%%%%%%%%

\section{Impact of measurements of the drift of the drift}

We now discuss the impact of using measurements of the drift of the drift. In this section we focus the discussion on the deceleration and jerk parameters (and the constraints in the corresponding two-dimensional plane), but the analysis to be discussed is still based on the full cosmographic series, up to and including the pop parameter, with these higher order parameters being marginalized.

We start by considering the case, already discussed in Sect. \ref{hods}, of using pairs of redshift drift measurements (at redshifts near each other) for calculating the drift of the drift through finite difference numerical derivatives, cf. Eqs. \ref{eq:dz1dz}--\ref{eq:Derr}. In doing this one may expect some information loss, especially at the higher redshifts, for the reasons already discussed in Sect. \ref{hods}. On the other hand, we also expect that the second derivative will have different sensitivities to the cosmographic parameters. The quantitative impact of the two effects can quantified by considering three separate cases. In our 'First' case, the 10 measurements of the first derivative data (i.e. of the canonical redshift drift) are used in the standard way. In the 'Second' case, the 10 measurements of the first derivative are converted into 9 measurements of the drift of the drift, as previously discussed. Finally, we have a 'Mixed' case, in which the three lowest redshift first derivative measurements  (i.e., up to and including $z=0.3$) yield two measurements of second derivative data, while beyond that one keeps the seven first derivative measurements. 

A further degree of freedom in our analysis stems from the choice of prior for the Hubble constant, and we consider three such choices: $\sigma_h=0$, $\sigma_h=0.05$ and $\sigma_h=0.1$, with the first being clearly optimistic while the latter is (arguably) pessimistic.

\begin{figure*}
\begin{center}
\includegraphics[width=0.32\columnwidth,keepaspectratio]{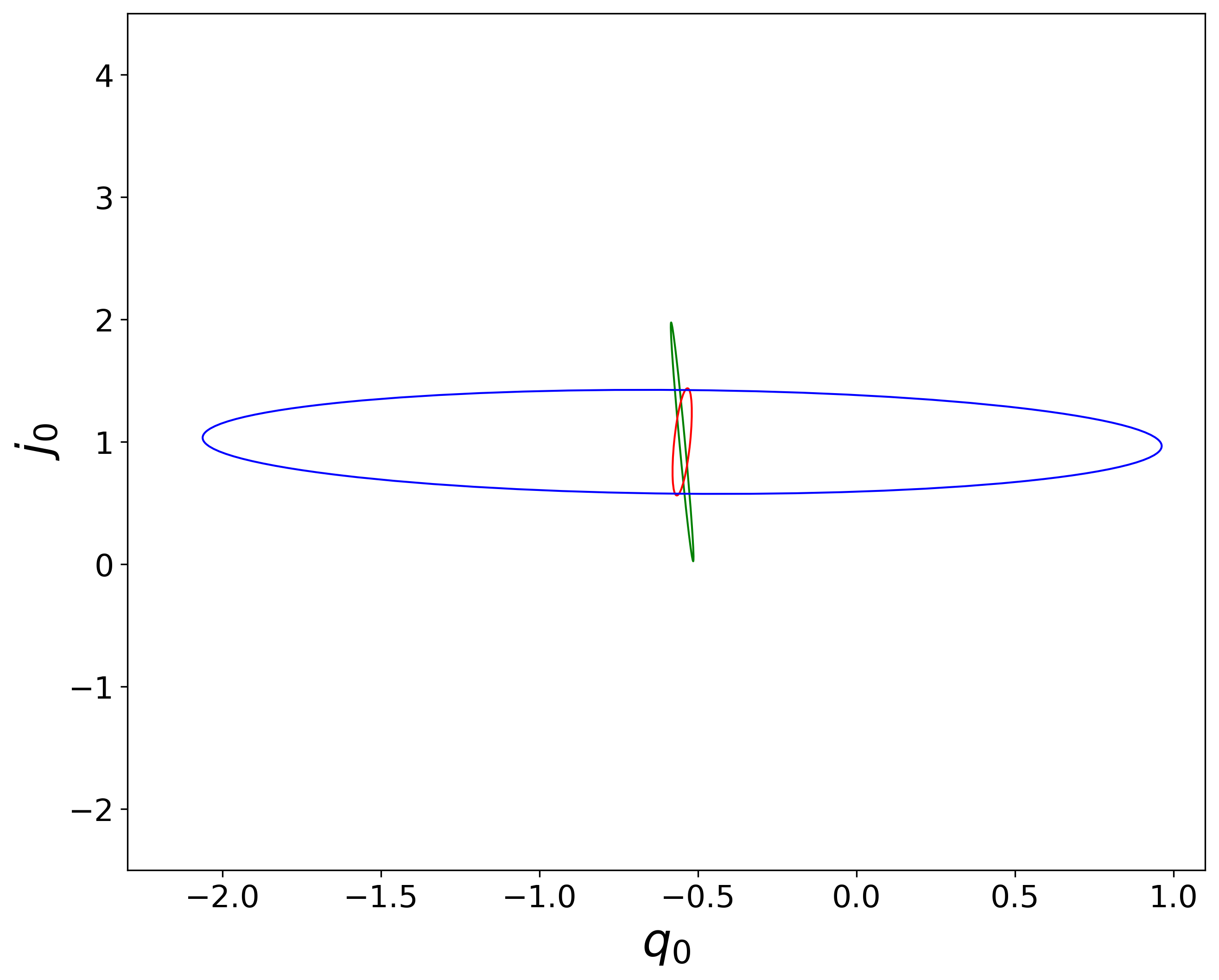}
\includegraphics[width=0.32\columnwidth,keepaspectratio]{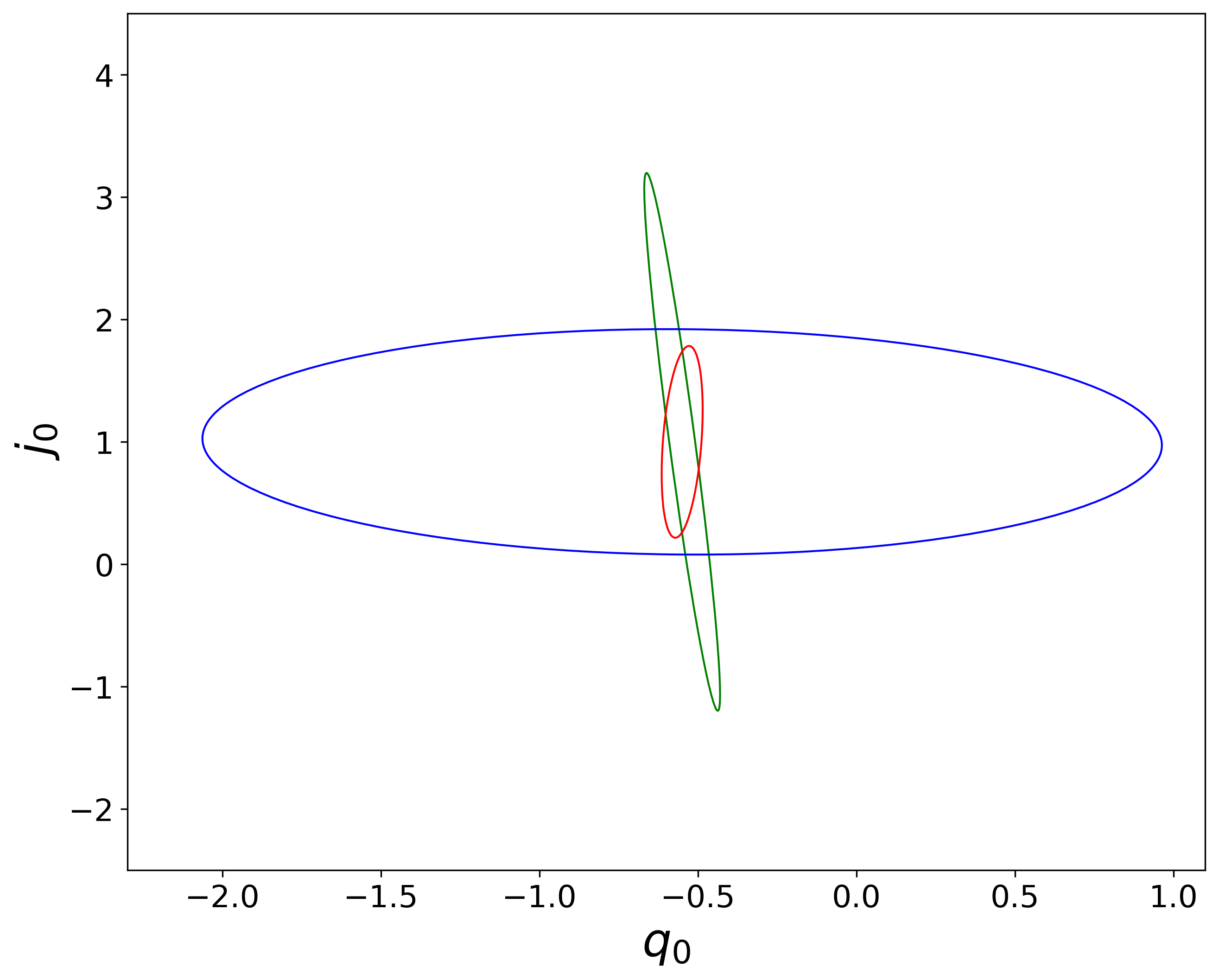}
\includegraphics[width=0.32\columnwidth,keepaspectratio]{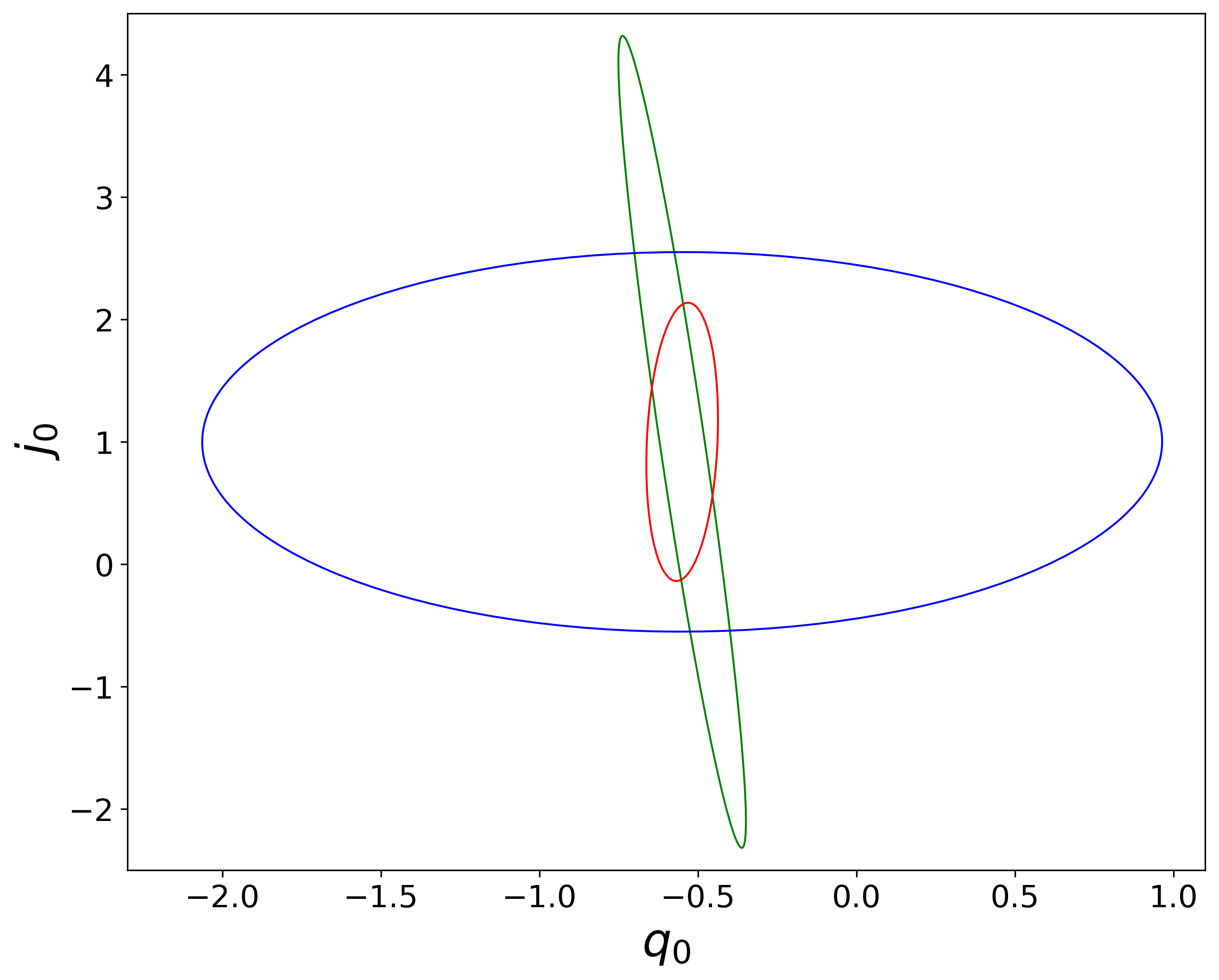}
\end{center}
\caption{One-sigma constraints of the $j_{\mathrm{0}}$-$q_{\mathrm{0}}$ plane, for scenarios in which some first derivative measurements are combined into a finite difference measurement of the second derivative. The first, mixed and second cases (discussed in the text) are shown in green, red and blue, respectively. The three panels correspond to the three choices of the prior on $h$, respectively $\sigma_h=0$, $\sigma_h=0.05$ and $\sigma_h=0.1$ from left to right. The axes are the same in all three panels, in order to facilitate a visual comparison.}
\label{fig3}
 \end{figure*}

Figure \ref{fig3} and Table \ref{tab3} show the results of this analysis. It is clear that for the Second case the information loss is substantial regardless of the prior on $h$, but this impacts only the posterior constraint on the deceleration parameter. On the other hand, the posterior constraint on the jerk parameter is always better in the Second case than in the First one, by a factor which is always larger than two. It is also worthy of note that for the second derivative the correlation coefficient between $q_{\mathrm{0}}$ and $j_{\mathrm{0}}$ is rather small, while for the first derivative the two parameters are strongly anti-correlated. Compared to the First approach, the Mixed one clearly improves the posterior constraints on both the deceleration and jerk parameters, though it may or may not improve the overall Figure of Merit (FoM), depending on the choice of prior for the Hubble constant, which impacts the correlation coefficient. This mixed approach seems therefore to be the optimal strategy. That being said, one should note the caveat that these conclusions do rely on a specific choice of fiducial model. A different underlying model, with a different redshift dependence of $D(z)$, may lead to quantitatively different conclusions. We leave the exploration of this possible model dependence for future work.

\begin{table}
    \centering
    \caption{Relevant results of the Fisher Matrix analysis for the comparison of the first and second derivatives of the redshift. The first row shows the Figure of Merit (rounded to the nearest integer), the second the correlation coefficients and the last two the one-sigma marginalised uncertainties. The First, Mixed and Second scenarios, referred to in the second row of the table header, are described in the main text. The left, middle and right sides of the table show the results for the three choices of priors for $h$, listed in the first row of the table header.}
    \label{tab3}
\begin{tabular}{l|ccc|ccc|ccc}
\multicolumn{1}{l|}{} & \multicolumn{3}{c|}{\textbf{$\sigma_h=0$}} & \multicolumn{3}{c|}{\textbf{$\sigma_h=0.05$}} & \multicolumn{3}{c|}{\textbf{$\sigma_h=0.1$}}\\
\textbf{Parameter}  &   \textbf{First} &  \textbf{Mixed} &  \textbf{Second} &   \textbf{First} &  \textbf{Mixed} &  \textbf{Second} &   \textbf{First} &  \textbf{Mixed} &  \textbf{Second}\\
\toprule
$\boldsymbol{\mathrm{FoM}(q_{\mathrm{0}}, j_{\mathrm{0}})}$ &   117 &     91 &       2 &      11 &     21 &       1 &       4 &      8 &       0 \\
\midrule
$\boldsymbol{\rho(q_{\mathrm{0}}, j_{\mathrm{0}})}$   &  -0.969 &  0.552 &  -0.080 &  -0.940 &  0.335 &  -0.030 &  -0.935 &  0.162 &   0.002 \\
\midrule
$\boldsymbol{\sigma(q_{\mathrm{0}})}$       &   0.024 &  0.020 &   0.995 &   0.079 &  0.043 &   0.996 &   0.132 &  0.074 &   0.996 \\
$\boldsymbol{\sigma(j_{\mathrm{0}})}$       &   0.641 &  0.288 &   0.280 &   1.445 &  0.515 &   0.606 &   2.182 &  0.748 &   1.020 \\
\bottomrule
\end{tabular}
\end{table}

\begin{figure*}
\begin{center}
\includegraphics[width=0.32\columnwidth,keepaspectratio]{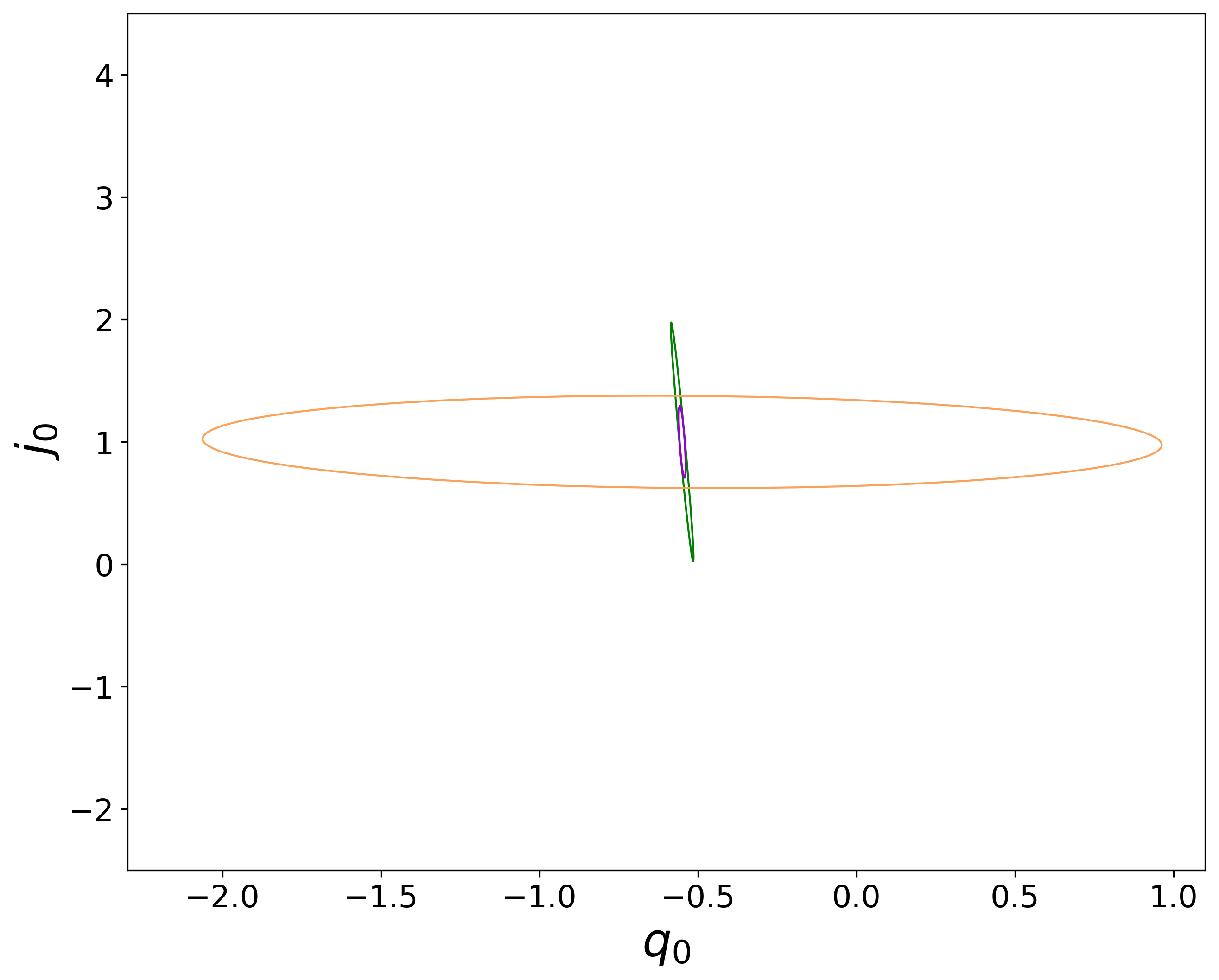}
\includegraphics[width=0.32\columnwidth,keepaspectratio]{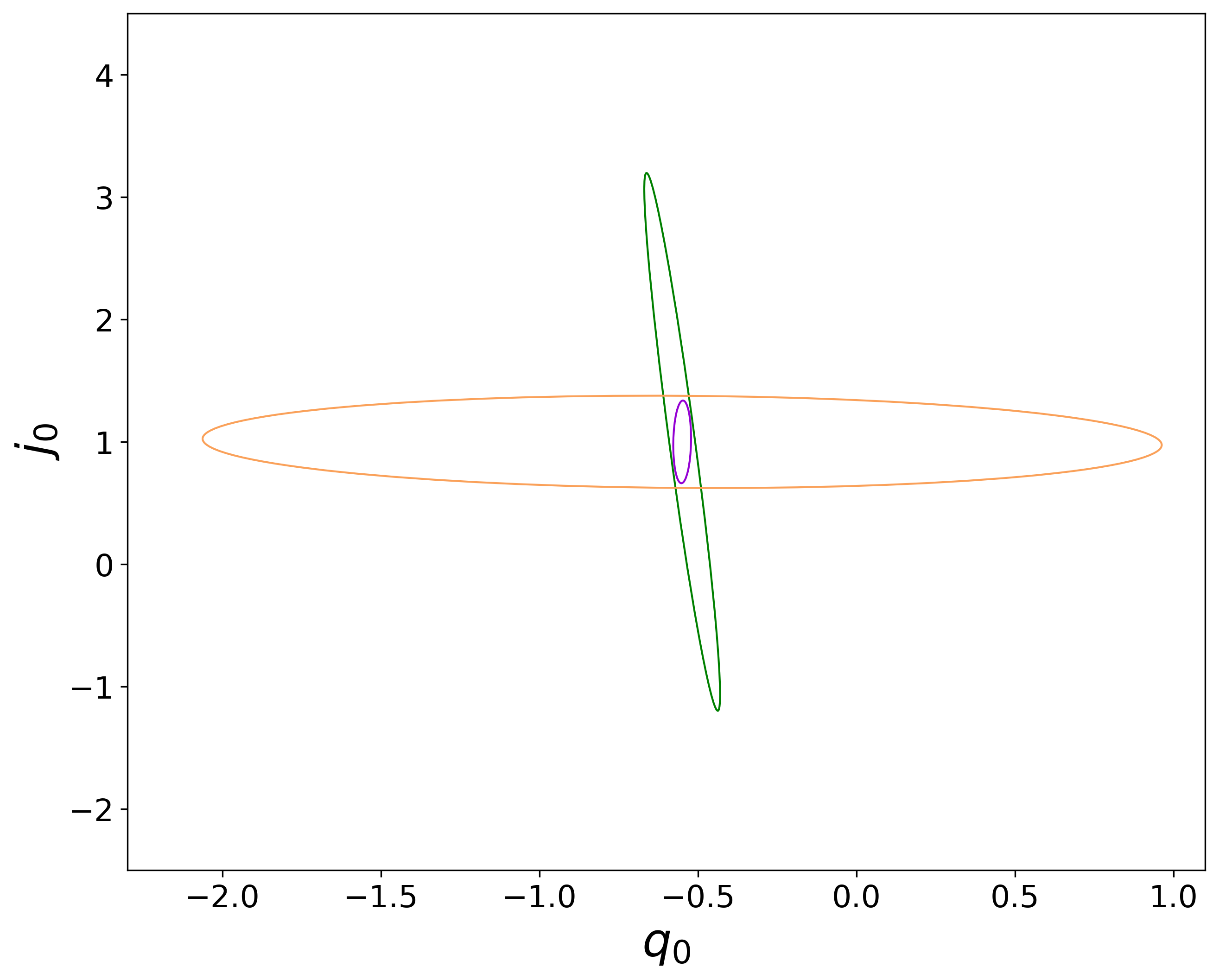}
\includegraphics[width=0.32\columnwidth,keepaspectratio]{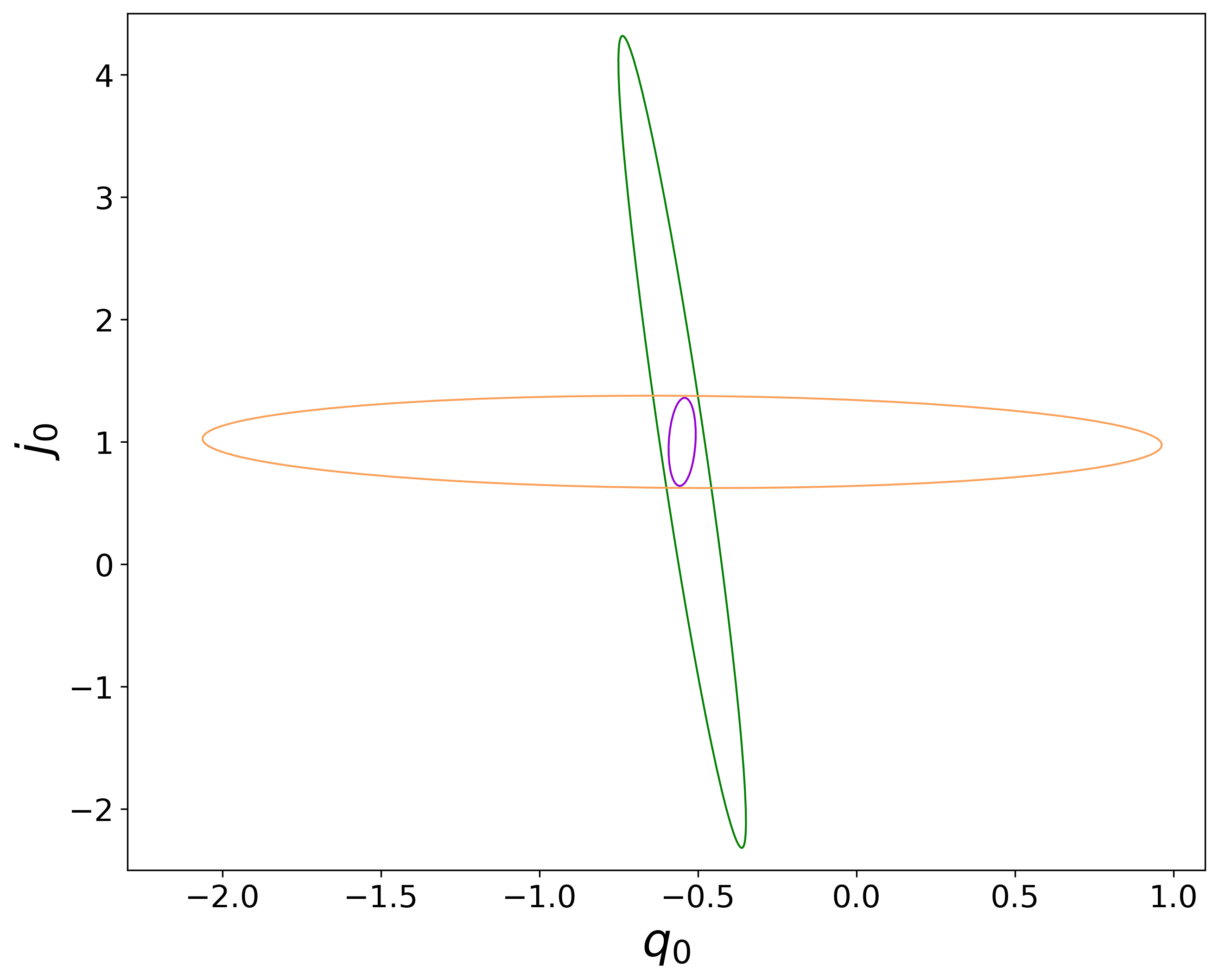}
\end{center}
\caption{One-sigma constraints of the $j_{\mathrm{0}}$-$q_{\mathrm{0}}$ plane, under the speculative assumption (further discussed in the text) of future direct measurements of the drift of the drift. The first, second and joint cases (also defined in the text) are shown in green, orange and purple, respectively. The three panels correspond to the three choices of the prior on $h$, respectively $\sigma_h=0$, $\sigma_h=0.05$ and $\sigma_h=0.1$ from left to right. The axes are the same in all three panels, in order to facilitate a visual comparison.}
\label{fig4}
 \end{figure*}
 
A second, fully speculative, possibility consists of assuming some future direct and separate (independent) measurements of $Z_{\mathrm{2}}(z)$, through an observational method which we leave unspecified, which would be added to the measurements of $Z_{\mathrm{1}}(z)$ which we have been discussing. Specifically, we assume three measurements at redshifts $z=0.1, 0.3, 0.5$, with relative uncertainties of $\sigma_{Z2}=10\%, 20\%, 30\%$ respectively. Note that, under these simplifying assumptions, there is no dependency on the prior on $h$. Table \ref{tab4} and Figure \ref{fig4} compare the constraints obtained from $Z_{\mathrm{1}}$ (already described in this section), with those from $Z_{\mathrm{2}}$ and the combination of the two datasets. In this case, as one may expect given the previous results, one sees significant gains in sensitivity. An important part of these gains stems from the fact that the strong anticorrelation between the two parameters, which occurs for the $Z_{\mathrm{1}}$ measurements, is alleviated by the addition of the almost orthogonal $Z_{\mathrm{2}}$ measurements.

\begin{table}
    \centering
    \caption{Relevant results of the Fisher Matrix analysis for the comparison of the measurements of $Z_{\mathrm{1}}$ and $Z_{\mathrm{2}}$. The first row shows the Figure of Merit (rounded to the nearest integer), the second the correlation coefficients and the last two the one-sigma marginalised uncertainties. The $Z_{\mathrm{1}}$, $Z_{\mathrm{2}}$ and Joint scenarios, referred to in the first row of the table header, are described in the main text. For the $Z_{\mathrm{1}}$ and joint cases, the results depend on the choice of the prior for $h$, listed in the first row of the table header.}
    \label{tab4}
\begin{tabular}{l|ccc|c|ccc}
\multicolumn{1}{l|}{} & \multicolumn{3}{c|}{\textbf{$Z_{\mathrm{1}}$}} & \multicolumn{1}{c|}{\textbf{$Z_{\mathrm{2}}$}} & \multicolumn{3}{c|}{\textbf{Joint}}\\
\textbf{Parameter}  &   \textbf{$\sigma_h=0$} &  \textbf{$\sigma_h=0.05$} &  \textbf{$\sigma_h=0.1$} & ($\sigma_h$ independent) &  \textbf{$\sigma_h=0$} &  \textbf{$\sigma_h=0.05$} &  \textbf{$\sigma_h=0.1$}\\
\toprule
$\boldsymbol{\mathrm{FoM}(q_{\mathrm{0}}, j_{\mathrm{0}})}$ & 117 & 11 & 4 & 2 & 430 & 106 & 66 \\
\midrule
$\boldsymbol{\rho(q_{\mathrm{0}}, j_{\mathrm{0}})}$ & -0.969 & -0.940 & -0.935 & -0.067 & -0.682 & 0.086 & 0.177 \\
\midrule
$\boldsymbol{\sigma(q_{\mathrm{0}})}$ & 0.024 & 0.079 & 0.132 & 0.995 & 0.007 & 0.018 & 0.028 \\
$\boldsymbol{\sigma(j_{\mathrm{0}})}$ & 0.641 & 1.445 & 2.182 & 0.248 & 0.192 & 0.222 & 0.237 \\
\bottomrule
\end{tabular}
\end{table}

%%%%%%%%%%%%%%%%%%%%%%%%%%%%%%%%%%%%%%%%%%%%%%%%%%%%%%%%%%%%%%%%%%%%%%%%%%%%%%%%%
%%%%%%%%%%%%%%%%%%%%%%%%%%%%%%%%%%%%%%%%%%%%%%%%%%%%%%%%%%%%%%%%%%%%%%%%%%%%%%%%%

\section{Constraining the deceleration and jerk}

Finally, we explore the impact of the choice of the order at which the cosmographic series is truncated on the derived constraints on the deceleration and jerk parameters, and additionally consider how this choice or truncation correlates with the redshift range of the available measurements. Towards this goal we now consider three scenarios: the full set of ten redshift drift measurements up to a maximum redshift $z=1.0$, and two subsets thereof, with maximum redshifts of $z=0.5$ and $z=0.3$. As before, we consider three possible priors on the Hubble constant, $\sigma_h=0$, $\sigma_h=0.05$ and $\sigma_h=0.1$.

Table \ref{tab5} shows the results of this analysis, for all combinations of the relevant variables (i.e., the truncation order of the series, prior on $h$, and redshift range of the measurements), while Figure \ref{fig5} depicts the results for the four 'extreme' cases of the truncation order ($n=5$ and $n=2$) and the Hubble parameter prior ($\sigma_h=0$ and $\sigma_h=0.1$), with all choices of redshift range depicted. In both the table and the figure a flat $\Lambda$CDM fiducial model, with $\Omega_{\mathrm{m}}=0.3$, $h=0.7$, $w_{\mathrm{0}}=-1$ and $w_{\mathrm{a}}=0$, has been assumed. For comparison, Table \ref{tab6} shows the analogous results for a freezing dynamical dark energy model, with $w_{\mathrm{0}}=-0.9$ and $w_{\mathrm{a}}=+0.3$, which was also used as a fiducial model in \citet{Marques}.

%%%%%%NEW TABLE
\begin{table}
    \centering
    \caption{Relevant results of the Fisher Matrix analysis for the comparison of the impact of the truncation order, Hubble constant prior and redshift range of the measurements on the constraints in the $q_{\mathrm{0}}$--$j_{\mathrm{0}}$ parameter space. A flat $\Lambda$CDM fiducial model, further defined in the text, has been assumed. The value of $n$ in the first column denotes the number of free cosmographic parameters (i.e., the order at which the series was truncated). Each series of rows shows the Figure of Merit (rounded to the nearest integer), the the correlation coefficient and the one-sigma marginalised uncertainties for each of the parameters. The subsequent columns show the results for the various combinations of the prior on $h$ and the maximum redshift of the data.}
    \label{tab5}
\begin{tabular}{l|l|ccc|ccc|ccc}
\multicolumn{1}{l|}{} & \multicolumn{1}{l|}{} & \multicolumn{3}{c|}{\textbf{$\sigma_h=0$}} & \multicolumn{3}{c|}{\textbf{$\sigma_h=0.05$}} & \multicolumn{3}{c|}{\textbf{$\sigma_h=0.1$}}\\
\textbf{}  & \textbf{Parameter}  &   \textbf{$z\le0.3$} &  \textbf{$z\le0.5$} &  \textbf{$z\le1.0$} &   \textbf{$z\le0.3$} &  \textbf{$z\le0.5$} &  \textbf{$z\le1.0$} &  \textbf{$z\le0.3$} &  \textbf{$z\le0.5$} &  \textbf{$z\le1.0$}\\
\toprule
\multirow{4}{*}{\textbf{n=5}}    & $\boldsymbol{\mathrm{FoM}(q_{\mathrm{0}}, j_{\mathrm{0}})}$ &   54 & 90 & 117 &       4 & 7 & 11 &           2 & 3 & 4 \\
                      & $\boldsymbol{\rho(q_{\mathrm{0}}, j_{\mathrm{0}})}$  &  -0.988 & -0.979 & -0.969 &      -0.944 & -0.954 & -0.940 &     -0.891 & -0.935 & -0.935 \\
                      & $\boldsymbol{\sigma(q_{\mathrm{0}})}$ &   0.042 & 0.029 & 0.024 &       0.112 & 0.098 & 0.079 &       0.154 & 0.152 & 0.132 \\
                      & $\boldsymbol{\sigma(j_{\mathrm{0}})}$   &  1.226 & 0.801 & 0.641 &       2.736 & 2.016 & 1.445 &       3.196 & 2.900 & 2.182 \\
\midrule
\multirow{4}{*}{\textbf{n=4}}    & $\boldsymbol{\mathrm{FoM}(q_{\mathrm{0}}, j_{\mathrm{0}})}$ &   37 & 54 & 245 &          4 & 5 & 18 &           2 & 2 & 6 \\
                      & $\boldsymbol{\rho(q_{\mathrm{0}}, j_{\mathrm{0}})}$  &  -0.983 & -0.986 & -0.956 &      -0.940 & -0.936 & -0.942 &      -0.890 & -0.908 & -0.938 \\
                      & $\boldsymbol{\sigma(q_{\mathrm{0}})}$ &   0.045 & 0.039 & 0.016 &       0.112 & 0.102 & 0.068 &       0.156 & 0.151 & 0.124 \\
                      & $\boldsymbol{\sigma(j_{\mathrm{0}})}$   &   1.422 & 1.203 & 0.368 &      2.736 & 2.365 & 1.063 &      3.200 & 2.969 & 1.822 \\
\midrule
\multirow{4}{*}{\textbf{n=3}}    & $\boldsymbol{\mathrm{FoM}(q_{\mathrm{0}}, j_{\mathrm{0}})}$ &   83 & 259 & 845 &         4 & 15 & 81 &          2 & 4 & 25  \\
                      & $\boldsymbol{\rho(q_{\mathrm{0}}, j_{\mathrm{0}})}$  &  -0.986 & -0.959 & -0.895 &      -0.959 & -0.948 & -0.933 &      -0.925 & -0.943 & -0.939 \\
                      & $\boldsymbol{\sigma(q_{\mathrm{0}})}$ &    0.034 & 0.017 & 0.009 &      0.123 & 0.076 & 0.041 &       0.177 & 0.141 & 0.077 \\
                      & $\boldsymbol{\sigma(j_{\mathrm{0}})}$   &  0.917 & 0.355 & 0.129 &       2.797 & 1.207 & 0.365 &       3.706 & 2.171 & 0.644 \\
\midrule
\multirow{4}{*}{\textbf{n=2}}    & $\boldsymbol{\mathrm{FoM}(q_{\mathrm{0}}, j_{\mathrm{0}})}$ &  788 & 1508 & 4050 &        42 & 128 & 652 &        11 & 37 & 236 \\
                      & $\boldsymbol{\rho(q_{\mathrm{0}}, j_{\mathrm{0}})}$  &  -0.917 & -0.865 & -0.805 &      -0.943 & -0.937 & -0.950 &      -0.944 & -0.942 & -0.963 \\
                      & $\boldsymbol{\sigma(q_{\mathrm{0}})}$ &   0.010 & 0.007 & 0.005 &       0.056 & 0.038 & 0.023 &       0.108 & 0.072 & 0.043 \\
                      & $\boldsymbol{\sigma(j_{\mathrm{0}})}$   &   0.138 & 0.078 & 0.033  &        0.562 & 0.259 & 0.092 &       1.081 & 0.483 & 0.160 \\
\bottomrule
\end{tabular}
\end{table}

\begin{table}
    \centering
    \caption{Same as Table \ref{tab5}, but for a fiducial freezing dark energy model, with  $w_{\mathrm{0}} = -0.9$ and $w_{\mathrm{a}} = 0.3$.}
    \label{tab6}
\begin{tabular}{l|l|ccc|ccc|ccc}
\multicolumn{1}{l|}{} & \multicolumn{1}{l|}{} & \multicolumn{3}{c|}{\textbf{$\sigma_h=0$}} & \multicolumn{3}{c|}{\textbf{$\sigma_h=0.05$}} & \multicolumn{3}{c|}{\textbf{$\sigma_h=0.1$}}\\
\textbf{}  & \textbf{Parameter}  &   \textbf{$z\le0.3$} &  \textbf{$z\le0.5$} &  \textbf{$z\le1.0$} &   \textbf{$z\le0.3$} &  \textbf{$z\le0.5$} &  \textbf{$z\le1.0$} &  \textbf{$z\le0.3$} &  \textbf{$z\le0.5$} &  \textbf{$z\le1.0$}\\
\toprule
\multirow{4}{*}{\textbf{n=5}}    & $\boldsymbol{\mathrm{FoM}(q_{\mathrm{0}}, j_{\mathrm{0}})}$ &   79 & 114 & 160 &       6 & 9 & 14 &           3 & 4 & 5 \\
                      & $\boldsymbol{\rho(q_{\mathrm{0}}, j_{\mathrm{0}})}$  &  -0.989 & -0.984 & -0.975 &      -0.955 & -0.962 & -0.952 &     -0.908 & -0.944 & -0.945 \\
                      & $\boldsymbol{\sigma(q_{\mathrm{0}})}$ &   0.042 & 0.033 & 0.011 &       0.100 & 0.090 & 0.072 &       0.136 & 0.135 & 0.118 \\
                      & $\boldsymbol{\sigma(j_{\mathrm{0}})}$   &  1.046 & 0.776 & 0.583 &       2.533 & 2.026 & 1.429 &       3.023 & 2.801 & 2.161 \\
\midrule
\multirow{4}{*}{\textbf{n=4}}    & $\boldsymbol{\mathrm{FoM}(q_{\mathrm{0}}, j_{\mathrm{0}})}$ &   43 & 84 & 532 &          5 & 7 & 40 &           2 & 3 & 12 \\
                      & $\boldsymbol{\rho(q_{\mathrm{0}}, j_{\mathrm{0}})}$  &  -0.985 & -0.987 & -0.946 &      -0.948 & -0.950 & -0.942 &      -0.905 & -0.916 & -0.942 \\
                      & $\boldsymbol{\sigma(q_{\mathrm{0}})}$ &   0.045 & 0.039 & 0.016 &       0.100 & 0.093 & 0.048 &       0.137 & 0.134 & 0.090 \\
                      & $\boldsymbol{\sigma(j_{\mathrm{0}})}$   &   1.392 & 0.994 & 0.226 &      2.542 & 2.288 & 0.671 &      3.017 & 2.884 & 1.189 \\
\midrule
\multirow{4}{*}{\textbf{n=3}}    & $\boldsymbol{\mathrm{FoM}(q_{\mathrm{0}}, j_{\mathrm{0}})}$ &   135 & 427 & 1269 &         7 & 24 & 118 &          2 & 7 & 37  \\
                      & $\boldsymbol{\rho(q_{\mathrm{0}}, j_{\mathrm{0}})}$  &  -0.986 & -0.957 & -0.898 &      -0.965 & -0.948 & -0.931 &      -0.944 & -0.946 & -0.936 \\
                      & $\boldsymbol{\sigma(q_{\mathrm{0}})}$ &    0.027 & 0.013 & 0.007 &      0.105 & 0.060 & 0.033 &       0.162 & 0.113 & 0.062 \\
                      & $\boldsymbol{\sigma(j_{\mathrm{0}})}$   &  0.711 & 0.272 & 0.107 &       2.402 & 0.934 & 0.303 &       3.487 & 1.720 & 0.534 \\
\midrule
\multirow{4}{*}{\textbf{n=2}}    & $\boldsymbol{\mathrm{FoM}(q_{\mathrm{0}}, j_{\mathrm{0}})}$ &  1384 & 2904 & 23717 &        74 & 257 & 4907 &        20 & 75 & 1913 \\
                      & $\boldsymbol{\rho(q_{\mathrm{0}}, j_{\mathrm{0}})}$  &  -0.913 & -0.859 & -0.923 &      -0.944 & -0.942 & -0.990 &      -0.946 & -0.948 & -0.993 \\
                      & $\boldsymbol{\sigma(q_{\mathrm{0}})}$ &   0.008 & 0.006 & 0.004 &       0.043 & 0.028 & 0.014 &       0.084 & 0.054 & 0.026 \\
                      & $\boldsymbol{\sigma(j_{\mathrm{0}})}$   &   0.101 & 0.053 & 0.013  &        0.416 & 0.179 & 0.043 &       0.802 & 0.336 & 0.077 \\
\bottomrule
\end{tabular}
\end{table}

The comparison of the various cases leads to several interesting aspects. First, while the results are qualitatively the same for both choices of fiducial model, they do depend quantitatively on this choice, by up to a factor of a few. Second, there is always a significant anticorrelation between the deceleration and jerk parameters, although its quantitative value depends on the other parametric choices. Third, the FoM increases as expected for larger datasets and smaller Hubble parameter prior uncertainties, and decreases if one keeps more terms in the cosmographic series. There is, however, one unexpected exception to the latter behaviour: in the case $n=4$ (corresponding to a series up to and including the crackle term), and for $z\le0.3$ and $z\le0.5$ (but not for $z\le1$) the FoM is smaller than in the corresponding $n=5$ cases. While this can presumably be due to the particular combinations of the cosmographic parameters in each of the terms of the series, we have not been able to identify the specific physical reason behind this behaviour.

Overall, one sees that such a dataset can lead to stringent constraints on both the deceleration and jerk parameters, although the order at which the series is truncated---which, on physical grounds, should be commensurate with the redshift range of the measurements---does have a significant impact on the result. In passing, we also note that these constraints are compatible with those of \citet{Second}, which obtained forecast constraints of $\sigma_{q0}\sim 0.006$ $\sigma_{j0}\sim 0.13$ from $z<0.3$ data, under somewhat different assumptions, e.g. a Planck-related prior for the Hubble constant. The trade-off between precision and accuracy of these constraints warrants additional work.

\begin{figure*}
\begin{center}
\includegraphics[width=0.45\columnwidth,keepaspectratio]{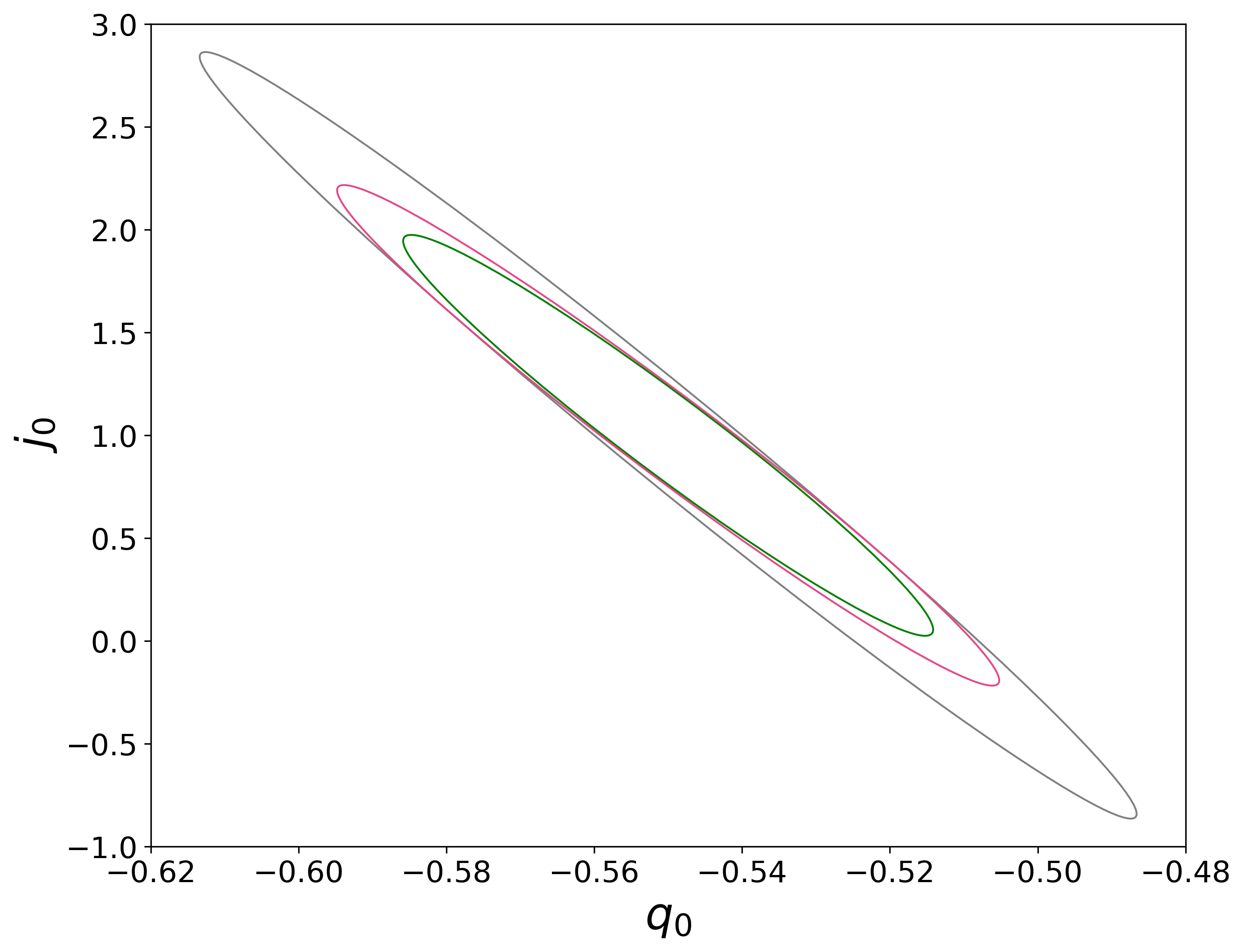}
\includegraphics[width=0.45\columnwidth,keepaspectratio]{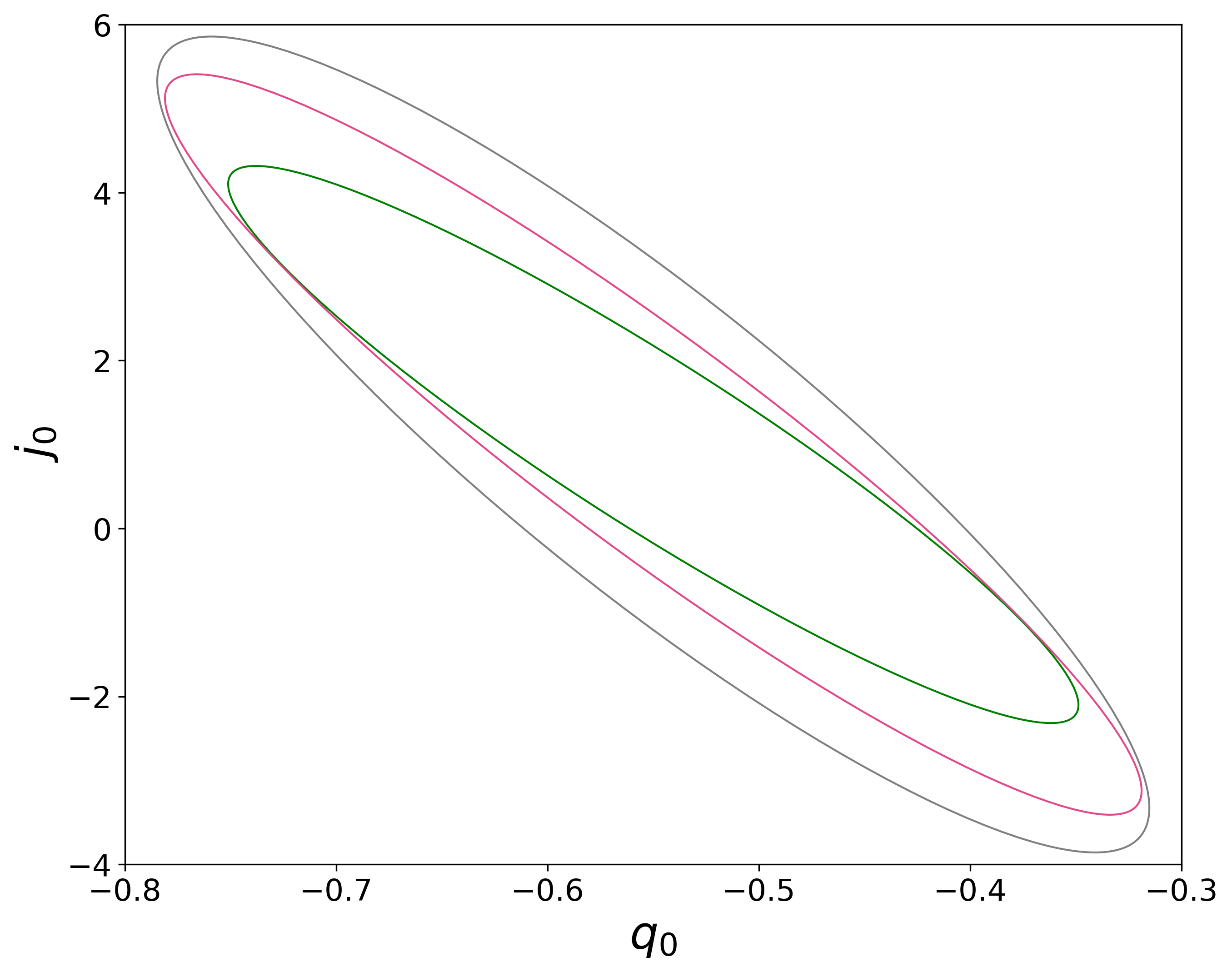}
\includegraphics[width=0.45\columnwidth,keepaspectratio]{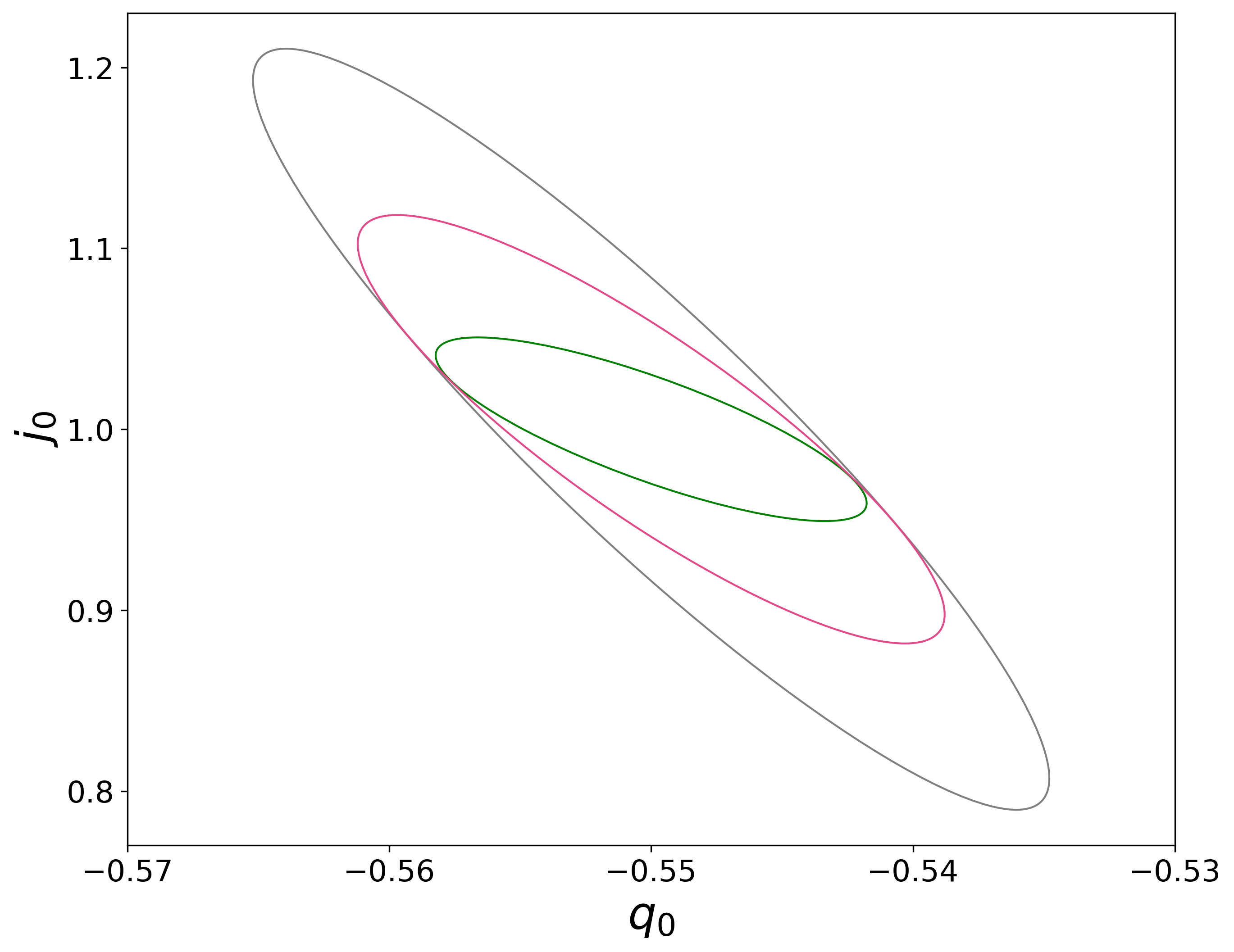}
\includegraphics[width=0.45\columnwidth,keepaspectratio]{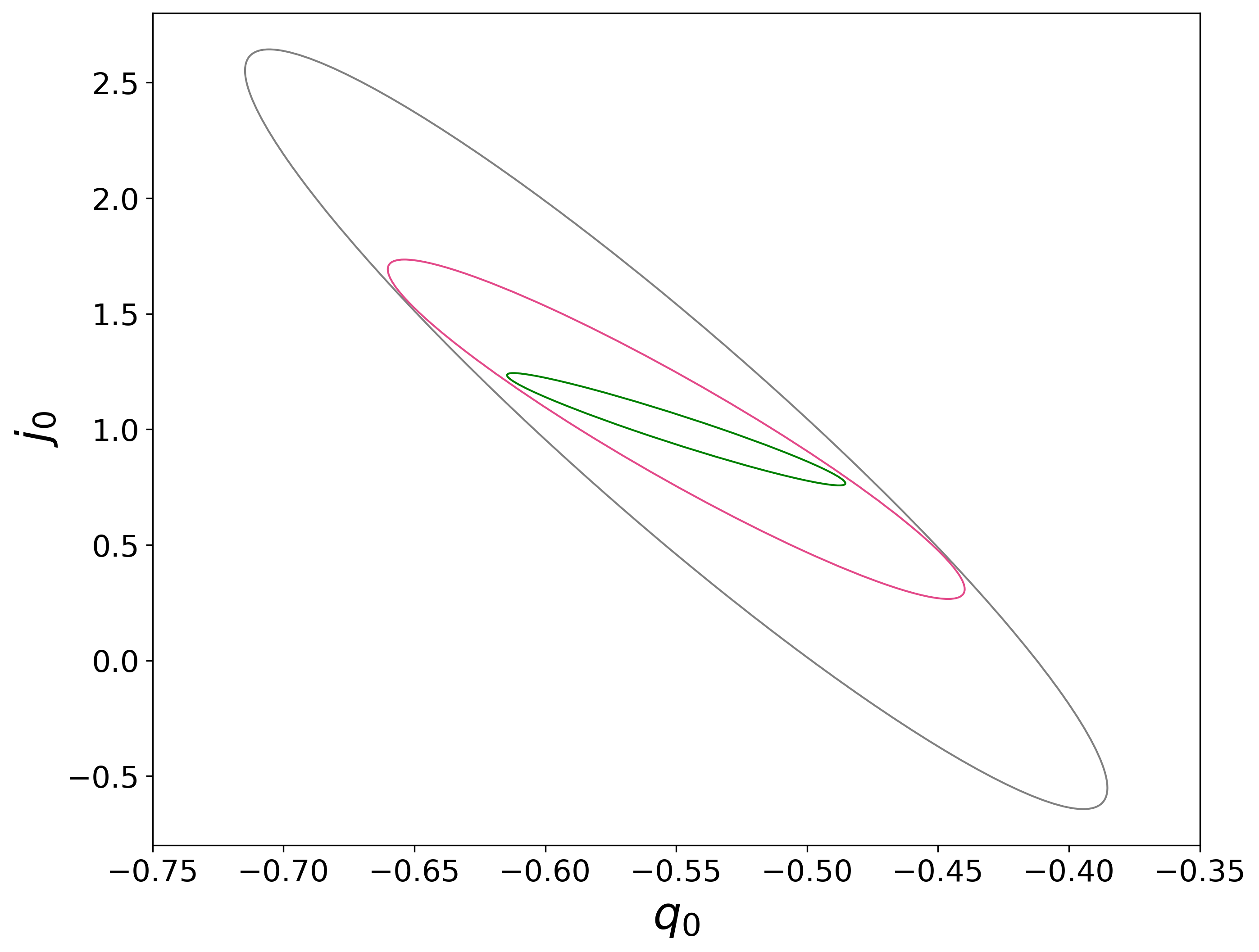}
\end{center}
\caption{One-sigma constraints in the $j_{\mathrm{0}}$-$q_{\mathrm{0}}$ plane, for a flat $\Lambda$CDM fiducial model. The data up to a maximum redshift of $z_{max}=0.3$, $z_{max}=0.5$ and $z_{max}=1.0$ are shown in grey, pink and green respectively. In the upper panels the cosmographic series has been truncated at $n=5$ and in the bottom ones $n=2$. The left panels have the prior $\sigma_h=0$ and the right panels have $\sigma_h=0.1$. Note that in this case each panel has different axis ranges.}
\label{fig5}
 \end{figure*}

%%%%%%NEW TABLE
\begin{table}
    \centering
    \caption{Relevant results of the Fisher Matrix analysis for the constraints from the redshift drift datasets discussed in the text on the $w_{\mathrm{0}}$--$w_{\mathrm{a}}$ parameter space for two CPL fiducial models. Each series of rows shows the Figure of Merit (rounded to the nearest integer), the the correlation coefficient and the the one-sigma marginalised uncertainties for each of the parameters. The subsequent columns show the results for the various combinations of the prior on $h$ and the maximum redshift of the data.}
    \label{tab7}
\begin{tabular}{l|l|ccc|ccc|ccc}
\multicolumn{1}{l|}{} & \multicolumn{1}{l|}{} & \multicolumn{3}{c|}{\textbf{$\sigma_h=0$}} & \multicolumn{3}{c|}{\textbf{$\sigma_h=0.05$}} & \multicolumn{3}{c|}{\textbf{$\sigma_h=0.1$}}\\
\textbf{}  & \textbf{Parameter}  &   \textbf{$z\le0.3$} &  \textbf{$z\le0.5$} &  \textbf{$z\le1.0$} &   \textbf{$z\le0.3$} &  \textbf{$z\le0.5$} &  \textbf{$z\le1.0$} &  \textbf{$z\le0.3$} &  \textbf{$z\le0.5$} &  \textbf{$z\le1.0$}\\
\toprule
                \multirow{4}{*}{\textbf{$\Lambda$CDM}}    & $\boldsymbol{\mathrm{FoM}(w_{\mathrm{0}}, w_{\mathrm{a}})}$ &  95 & 139 & 252 &        48 & 62 & 110 &        44 & 50 & 77 \\
                      & $\boldsymbol{\rho(w_{\mathrm{0}}, w_{\mathrm{a}})}$  &  -0.624 & -0.803 & -0.905 &      -0.226 & -0.433 & -0.761 &      -0.156 & -0.287 & -0.609 \\
                      & $\boldsymbol{\sigma(w_{\mathrm{0}})}$ &   0.038 & 0.034 & 0.024 &        0.054 & 0.051 & 0.040 &       0.062 & 0.062 & 0.056 \\
                      & $\boldsymbol{\sigma(w_{\mathrm{a}})}$   &   0.230 & 0.214 & 0.207  &        0.352 & 0.309 & 0.285 &       0.376 & 0.349 & 0.316 \\
\midrule                      
\multirow{4}{*}{\textbf{Freezing}}    & $\boldsymbol{\mathrm{FoM}(w_{\mathrm{0}}, w_{\mathrm{a}})}$ &  118 & 183 & 352 &        50 & 71 & 139 &        44 & 53 & 91 \\
                      & $\boldsymbol{\rho(w_{\mathrm{0}}, w_{\mathrm{a}})}$  &  -0.614 & -0.805 & -0.934 &      -0.221 & -0.432 & -0.798 &      -0.154 & -0.288 & -0.647 \\
                      & $\boldsymbol{\sigma(w_{\mathrm{0}})}$ &   0.035 & 0.033 & 0.027 &        0.048 & 0.045 & 0.035 &       0.055 & 0.055 & 0.048 \\
                      & $\boldsymbol{\sigma(w_{\mathrm{a}})}$   &   0.172 & 0.151 & 0.147  &        0.325 & 0.254 & 0.205 &       0.365 & 0.317 & 0.245 \\
\bottomrule
\end{tabular}
\end{table}

Finally, for comparison purposes, we step back from the cosmographic approach and return to our particular choice of the CPL fiducial model. Table \ref{tab7} shows, for the various assumptions on redshift drift measurement data under consideration in this section, and for both the $\Lambda$CDM and the freezing fiducial models, the derived constraints in the $(w_{\mathrm{0}},w_{\mathrm{a}})$ plane, with the rest of the CPL model parameters marginalized as usual. Again, we see that the quantitative results are affected by the choice of fiducial model and the prior on $H_{\mathrm{0}}$, although here the differences are smaller than for the cosmographic approach. Unsurprisingly, the same can be said for the redshift range of the redshift drift measurements: here going from $z\le0.3$ to $z\le1$ improves the FoM by at most a factor of three, while in the cosmographic approach the improvement often exceeds a factor of ten. This shows the importance of combining SKAO redshift drift measurements with those obtained at higher redshifts by the ELT, as discussed in \citet{Rocha}. On the other hand, for this phenomenological CPL parameterization and other models, additional gains in sensitivity can be obtained by combining redshift drift measurements with traditional cosmological probes such as Type Ia supernovae and the cosmic microwave background \citep{MG16}.

%%%%%%%%%%%%%%%%%%%%%%%%%%%%%%%%%%%%%%%%%%%%%%%%%%%%%%%%%%%%%%%%%%%%%%%%%%%%%%

\section{Conclusions}

We have relied on Fisher Matrix Analysis techniques to discuss the potential cosmological impact of SKAO measurements of the redshift drift. These measurements complement those of the ELT, probing the low redshift range ($z<1$), and can therefore enable us to watch the acceleration phase of the Universe in real time. We should note that at present there are no detailed simulations of the expected SKAO performance that are directly relevant for its redshift drift measurements, comparable to what has been done for the ELT \citep{Liske,Cooke,Qubrics}. Thus, our analysis took as starting point the comparatively simpler estimates of \citet{Klockner}, as was the case in other recent works \citep{Alves,Rocha}.

Having established the reliability of our Fisher Matrix approach, by comparison to the Markov Chain Monte Carlo analysis in \citet{Rocha}, and also that of our choices of priors for the parameters in our cosmographic series approach, we have obtained forecasts under several assumptions on datasets containing measurements of the redshift drift and its derivative. These various assumptions aim to encompass, in a simplified way, the range of possible theoretical scenarios (e.g., through different choices of fiducial model) but also of possible observational scenarios and trade-offs (e.g., through different choices on the redshift range of the available measurements and on the Hubble parameter prior). We have not considered the experiment time and on-sky observation time as variables in our analysis, since the approach, following from \citet{Klockner}, of expressing the measurement uncertainties as relative (percent level) errors renders them unnecessary.

Our analysis has predominantly focused on the cosmographic approach, and especially on the constraints on the deceleration and jerk parameters. As we have pointed out, in the context of the redshift drift, the Hubble constant plays a role that is physically different from that of the other coefficients in the series, and it is best seen as an overall normalization (multiplicative) factor. The cosmographic approach does have the significant advantage of model independence, and in this regard, we also recall the earlier work of \citet{Neben}, which argues that the redshift drift is an optimal way to measure the deceleration parameter because it allows its direct measurement with both accuracy and precision. Nevertheless, one faces the question of the truncation of the cosmographic series---keeping additional terms will generally mean more accuracy but less precision. We have not addressed this issue in detail, mainly because to reliably do so one would need more realistic SKAO-like simulations. Beyond $z=1$ (a redshift range beyond the reach of SKAO redshift drift measurements, but relevant for the combination of these measurements with those from the ELT) one also faces the issue of convergence of the cosmographic series, but this can be addressed by a series expansion using variables related to the redshift rather than the redshift itself, as discussed in \citet{Rocha}.

Overall, and to the extent that our assumptions on the SKAO performance are realistic, our analysis shows that in a reasonable amount of observing time it can provide a set of redshift drift measurements, including the first derivative of the drift, whose constraining power will be competitive with that of traditional  probes of the low-redshift Universe. Given the SKAO measurement method, which will rely on stacking information from a significant area of the sky, it may also enable other possibilities, such as a cross-correlation with galaxy number counts \citep{Bessa}. And, last but not least, one must not forget the conceptually fundamental point that a direct measurement of a positive redshift drift demonstrates a violation of the strong energy condition, and therefore implies the presence of dark energy.

\section*{Acknowledgements}

This work was financed by Portuguese funds through FCT - Funda\c c\~ao para a Ci\^encia e a Tecnologia in the framework of the project 2022.04048.PTDC (Phi in the Sky, DOI 10.54499/2022.04048.PTDC). CMJM is supported by an FCT fellowship, grant number 2023.03984.BD. CJAPM also acknowledges FCT and POCH/FSE (EC) support through Investigador FCT Contract 2021.01214.CEECIND/CP1658/CT0001. We acknowledge useful discussions on the topic of this work with Catarina Alves, Sofia Fernandes, Axel Lapel and Mariana Melo e Sousa.

{\noindent\bf Data availability:} This work uses simulated data, generated as detailed in the text. The codes used herein are based on the publicly available FRIDDA code \citep{FRIDDA}, and can be made available to interested colleagues on request to the first author.

\bibliographystyle{mnras}
\bibliography{drift} % if your bibtex file is called example.bib
%%%%%%%%%%%%%%%%%%%%%%%%%%%%%%%%%%%%%%%%%%%%%%%%%%
% Don't change these lines
\bsp	% typesetting comment
\label{lastpage}
\end{document}